\documentclass[twocolumn,english,prb,aps,showpacs,floatfix]{revtex4}
\usepackage{lmodern}
\usepackage[T1]{fontenc}
\usepackage[latin9]{inputenc}
\usepackage{color}
\usepackage{amsmath}
\usepackage{graphicx}
\usepackage{amssymb}
\usepackage{esint}

\makeatletter
\@ifundefined{textcolor}{}
{%
 \definecolor{BLACK}{gray}{0}
 \definecolor{WHITE}{gray}{1}
 \definecolor{RED}{rgb}{1,0,0}
 \definecolor{GREEN}{rgb}{0,1,0}
 \definecolor{BLUE}{rgb}{0,0,1}
 \definecolor{CYAN}{cmyk}{1,0,0,0}
 \definecolor{MAGENTA}{cmyk}{0,1,0,0}
 \definecolor{YELLOW}{cmyk}{0,0,1,0}
 }

\usepackage{amsxtra}\usepackage{bm}\usepackage{bbm}

\makeatother

\usepackage{babel}

\makeatother

\usepackage{babel}

\begin{document}

\title{Orbital symmetry fingerprints for magnetic adatoms in graphene}

\author{Bruno Uchoa$^{1,2}$, Ling Yang$^{3}$, S.-W. Tsai$^{3}$, N.~M.~R.
Peres$^{4}$, and A.~H. Castro Neto$^{5,6}$}

\affiliation{$^{1}$Department of Physics and Astronomy, University of Oklahoma,
440 W. Brooks St., Norman, OK 73019}

\affiliation{$^{2}$Department of Physics$\mbox{,}$ University of Illinois at
Urbana-Champaign, 1110 W. Green St, Urbana, IL, 61801, USA}

\affiliation{$^{3}$Department of Physics and Astronomy, University of California,
Riverside, CA, 92521,USA}

\affiliation{$^{4}$Centro de Física e Departamento de Física, Universidade do
Minho, P-4710-057, Braga, Portugal}

\affiliation{$^{5}$Graphene Research Centre, and Department of Physics, National
University of Singapore, 2 Science Drive 3, Singapore 117542}

\affiliation{$^{6}$Department of Physics, Boston University, 590 Commonwealth
Avenue, Boston, MA 02215, USA}

\date{\today}
\begin{abstract}
In this paper, we describe the formation of local resonances in graphene
in the presence of magnetic adatoms containing localized orbitals
of arbitrary symmetry, corresponding to any given angular momentum
state. We show that quantum interference effects which are naturally
inbuilt in the honeycomb lattice in combination with the specific
orbital symmetry of the localized state lead to the formation of fingerprints
in differential conductance curves.\textcolor{black}{{} In the presence
of Jahn-Teller distortion effects, which lift the orbital degeneracy
of the adatoms, the orbital symmetries can lead to distinctive signatures
in the local density of states.}\textcolor{blue}{{} }We show that those
effects allow scanning tunneling probes to characterize adatoms and
defects in graphene. 
\end{abstract}

\pacs{73.20.Hb,71.55.Ht,73.20.-r}

\maketitle

\section{Introduction}

Graphene is a single atomic layer of graphite whose low energy quasiparticles
behave as massless Dirac fermions\cite{novo3,zhang,neto}. As an open
surface, graphene offers a solid playground for the detection and
local manipulation of quantum states with scanning tunneling (STM)
probes. This perspective is particularly promising for adatoms, which
can be dragged with atomic precision\cite{eigler} and can have their
magnetic state monitored and controlled with the application of an
external gate voltage\cite{uchoa,sengupta}. There has been substantial
progress in the quality of the STM experiments in graphene in the
last few years\cite{stolyarova,Rutter,Brar,Ishigami,zhang2,Geringer,Li,Xue,Deshpande,Brihuega}.
Recent experiments reported the observation of Landau levels spontaneously
generated by strain on the top of nanobubbles in graphene\cite{Levy},
and the observation of charge polarization effects around a Co adatom\cite{Brar2}.

Although the microscopic theory of STM is well understood in metallic
hosts\cite{Tersoff,plihal}, in graphene the sublattice quantum numbers
play a role in the interference effects that drive the emergence of
Fano resonances\cite{fano} nearby adatoms, in the presence of an
STM tip. In particular, for adatoms that sit at the center of the
honeycomb hexagon ($H$ site), destructive interference between the
different electronic paths of hybridization with the two sublattices
may give rise to a suppression of the Fano resonance of the localized
state\cite{Uchoa2,sengupta2,Wehling}, and also change the scattering
rate of the localized electrons due to the presence of the fermionic
bath\cite{Uchoa2}. In general, the broadening of a localized state,
magnetic or not, is expected to scale as $\Delta(\omega)\propto|\omega|^{r}$,
where $r$ is the scaling dimension of the DOS of the host material,
which in graphene is $r=1$ ($r=0$ for metals). In graphene, nevertheless,
localized orbitals located either in substitutional impurity sites
($S$ sites) or in $H$ sites and which also preserve the $C_{3v}$
point group symmetry of each sublattice are effectively damped at
low energies by a fermionic bath with $r=3$\cite{Uchoa2,Uchoa3},
due to quantum interference effects. This effect suggests that the
local density of states (LDOS) can by quite susceptible to the orbital
symmetry of the localized state, allowing STM probes to characterize
adatoms and defects in graphene. 

\textcolor{black}{In this work, we describe in detail the effect of
the localized orbital symmetry in the emergence of local magnetic
resonances near the adatoms with inner shell electrons. We discuss
the emergence of non-trivial particle-hole asymmetries in the energy
dependence of the level broadening $\Delta(\omega)$, depending on
the particular symmetry and position of the localized state in the
lattice. We also describe the way the differential conductance curves
reflect the orbital symmetry of spin polarized states. }

\textcolor{black}{In real crystals, where the adatoms are randomly
distributed, local lattice distortions created by the adatoms\cite{Liu2}
can displace them from high symmetry positions in the crystal. In
the presence of local Jahn-Teller distortions that lift orbital degeneracies,
we also show that the adatoms can induce distinctive signatures of
the individual orbital symmetries directly in the LDOS of graphene,
which can be measured with local energy resolved spectroscopy experiments.
This effect is not present in ordinary host metals. We will address
the limiting situation where the charge of the orbitals in a given
irreducible representation is strongly polarized. In graphene, where
orbital degeneracies appear in the form of doublet states, this limit
can be physically described by adatoms with total spin $1/2$, when
one of the orbitals in the doublet is half filled (spin polarized)
and the other empty. In this regime, which will be assumed for most
of the paper, the problem can be described by an effective }\textcolor{black}{\emph{single}}\textcolor{black}{{}
orbital Hamiltonian. }In the second part of the paper, we address
the theory of scanning tunneling spectroscopy developed in ref. \cite{Uchoa2}
for the case of $s$-wave orbitals, and generalize it to describe
higher angular momentum states in the case of interest, for adatoms
with total spin near $1/2$. 

The outline of the paper is as follows: in section II, we describe
the generic zero dimensional Hamiltonian of an adatom in graphene;
in section III we briefly describe the role of the orbital symmetry
into the formation of local magnetic moments and we show the manifestation
of those orbital symmetries in the LDOS, whenever the adatom hybridizes
with two or more carbon atoms. We address in particular the appearance
of particle-hole asymmetries in the level broadening observed in ab
initio calculations. In section IV we address the STM tip effects
in the LDOS and we compute the differential conductance accounting
for the symmetry of the localized orbitals and their position with
respect to the sublattices. Finally, in section V we present our conclusions.

\section{Hamiltonian}

The Hamiltonian of a magnetic adatom in graphene is described by a
sum of four terms, \begin{equation}
H=H_{g}+H_{f}+H_{V}+H_{U}\,,\label{eq:H1}\end{equation}
 where\begin{equation}
H_{g}=-t\sum_{\langle ij\rangle}a_{\sigma}^{\dagger}(\mathbf{R}_{i})b_{\sigma}(\mathbf{R}_{j})+h.c\label{eq:Hg}\end{equation}
 is the graphene Hamiltonian in tight-binding, with $t\sim2.8$eV
the hoping energy between nearest neighbors sites, $a$ ($b$) is
a fermionic annihilation operator in the $A$ ($B$) sublattice, with
$\sigma=\uparrow,\downarrow$ indexing the spin. In momentum space,
\begin{equation}
H_{g}=-t\sum_{\mathbf{p}\sigma}\left(\phi_{\mathbf{p}}a_{\mathbf{p}\sigma}^{\dagger}b_{\mathbf{p}\sigma}+h.c.\right)\label{graphene}\end{equation}
 \\
 where $\phi_{\mathbf{k}}=\sum_{i=1}^{3}\mbox{e}^{i\mathbf{k}\cdot\mathbf{a}_{i}}$
and $\mathbf{a}_{1}=\hat{x}$, $\mathbf{a}_{2}=-\hat{x}/2+\sqrt{3}\hat{y}/2$,
and $\mathbf{a}_{3}=-\hat{x}/2-\sqrt{3}\hat{y}/2$ are the lattice
nearest neighbor vectors. The second term, \begin{equation}
H_{f}=\sum_{\sigma}\sum_{m}\epsilon_{0}\, f_{m,\sigma}^{\dagger}f_{m,\sigma}\label{Hf}\end{equation}
 is the Hamiltonian of the localized level with energy $\epsilon_{0}$
measured from the Dirac point, with $m$ the angular momentum projection
indexing the different degenerate orbitals contained in a given irreducible
representation with angular momentum $l$ (for instance, the doublets
$d_{xy}$, $d_{x^{2}-y^{2}}$, with $l=2$ and $m=\pm l$). In graphene,
due to the three fold rotational symmetry of each sublattice, the
crystalline filed anisotropy lifts the degeneracy of the orbitals
for different values of $|m|\leq l$, leaving pairs of degenerate
states (doublets) with angular momentum projections $\pm|m|$ with
$m\neq0$. 

\begin{figure}
\begin{centering}
\includegraphics[scale=0.27]{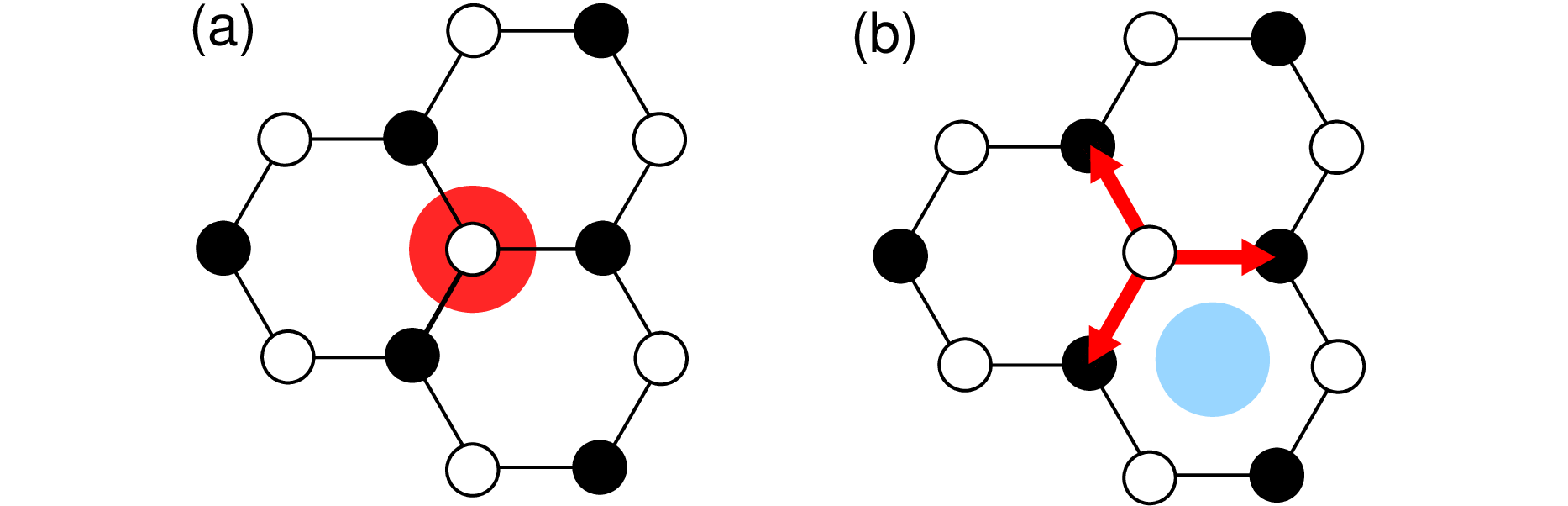} 
\par\end{centering}

\caption{{\small Honeycomb lattice in graphene, with two distinct sublattices
(black and white circles). a) Adatom siting on top of a carbon atom
on sublattice $A$, and b) sitting in the center of the honeycomb
hexagon. Red arrows:} {\small nearest neighbor vectors.\label{fig:Honeycomb-lattice-in}}}
\end{figure}

The third term in Eq.(\ref{eq:H1}) gives the hybridization Hamiltonian.
When the adatoms sit on top of a carbon atom as in the case of H and
F atoms, and also simple molecules\cite{Leenaerts} such as $\mbox{NO}_{2}$,
\[
H_{V}=V\sum_{\sigma,m}a_{\sigma}^{\dagger}(0)f_{m,\sigma}+h.c.\,,\]
 for adsorption on a given site, say on sublattice $A$. Adatoms such
as transition metals may instead strongly prefer to sit in the hollow
site \cite{Chan} at the center of the hexagon in the honeycomb lattice
(see Fig. 1). In that case, the level is coupled to the graphene bath
though the hybridization Hamiltonian\cite{Uchoa2} \begin{equation}
H_{V}=\sum_{\sigma,m}\sum_{i=1}^{3}\left[V_{a,i}^{(m)}a_{\sigma}^{\dagger}(\mathbf{a}_{i})+V_{b,i}^{(m)}b_{\sigma}^{\dagger}(-\mathbf{a}_{i})\right]f_{m,\sigma}+h.c.\label{eq:Hv}\end{equation}
 where $V_{x,i}$ ($x=a,\, b$) are the hybridization amplitudes of
the adatom with each of the nearest neighbors carbon atoms, which
are set by the orbital symmetry of the localized state. In momentum
space representation, this Hamiltonian can be written as\cite{Uchoa2,Uchoa3}
\begin{equation}
H_{V}=\frac{1}{\sqrt{N}}\sum_{m,\mathbf{p}\sigma}\left(V_{b,\mathbf{p}}^{(m)}b_{\mathbf{p}\sigma}^{\dagger}+V_{a,\mathbf{p}}^{(m)}a_{\mathbf{p}}^{\dagger}\right)f_{m,\sigma}+h.c.\,,\label{Hv2}\end{equation}
 where \begin{eqnarray}
V_{b,\mathbf{p}}^{(m)} & = & \sum_{\langle j\rangle}V_{b,j}^{(m)}\mbox{e}^{i\mathbf{p}\cdot\mathbf{a}_{j}}\label{eq:Vb}\\
V_{a,\mathbf{p}}^{(m)} & = & \sum_{\langle j\rangle}V_{a,j}^{(m)}\mbox{e}^{-i\mathbf{p}\cdot\mathbf{a}_{j}}\,,\label{eq:Va}\end{eqnarray}
 with $\langle j\rangle$ representing summation over the hybridization
amplitudes of the adatom with the nearest neighbor carbon atoms on
a given sublattice, and $N$ is the number of lattice sites in the
extended unit cell of the adatom. The discrete sum over momenta can
be interchanged by a continuous integration, $\frac{1}{N}\sum_{\mathbf{k}}\longrightarrow A\int\mbox{d}\mathbf{k}\,,$
where $A=2/D^{2}$, where $D\approx7$eV is the bandwidth. For notational
reasons, we will set $N=1$ from now on.

\begin{figure}[t]
\begin{centering}
\includegraphics[scale=0.23]{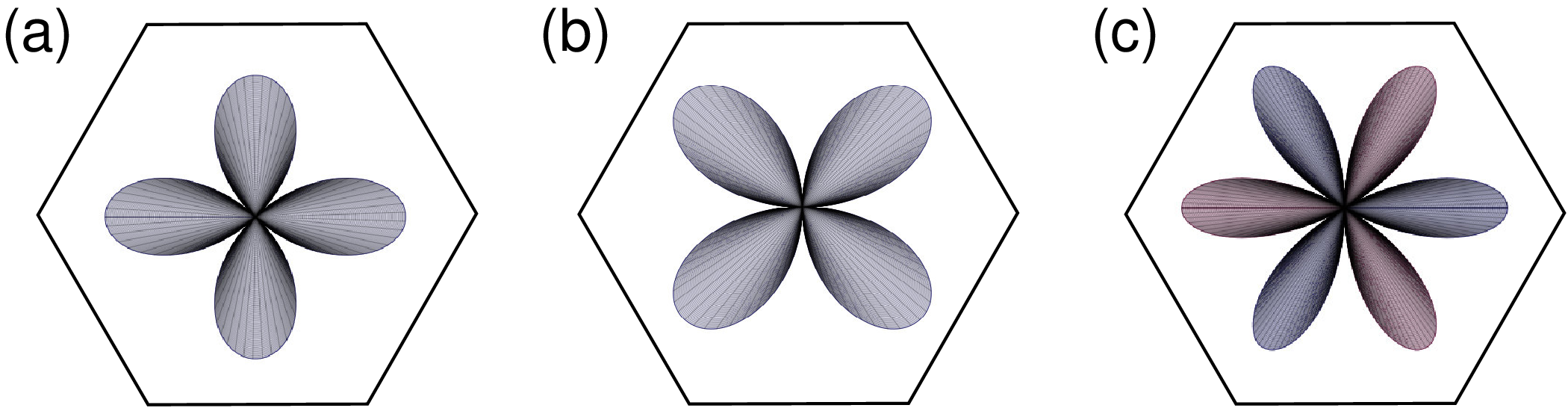} 
\par\end{centering}

\caption{{\small Illustration of $d$-wave and $f$-wave localized orbitals
for an adatom that sits at the center of the honeycomb lattice hexagon
(H site). a) $d_{x^{2}-y^{2}}$ orbital, that corresponds to hybridization
amplitudes $V_{x,1}=V$ and $V_{x,2}=V_{x,3}=-V/2$ with the six nearest
neighbor carbon atoms on sublattices $x=a,\, b$, at the vertexes
of the hexagon {[}see Eq. (\ref{eq:Vb}), (\ref{eq:Va}){]}. b) $d_{xy}$
orbital corresponding to $V_{x,1}=0$ and $V_{x,2}=-V_{x,3}=V$. c)
$f_{x(x^{2}-3y^{2})}$ orbital, with hybridization amplitudes $V_{a,i}=-V_{b,i}=V$.
Adatoms on top carbon sites, and adatoms in $H$ or $S$ sites with
$p$-wave, $d$-wave and out of plane $f$-wave orbitals explicitly
break the $C_{3v}$ sublattice symmetry in graphene (type $I$ orbitals).
$s$-wave and in-plane $f$-wave orbitals sitting on $H$ or $S$
sites are $C_{3v}$ invariant (type $II$ orbitals, see text) \label{fig:Illustration-of-d-wave}}}
\end{figure}

For adatoms that sit on top of the carbon atoms, (such as hydrogen),
$V_{a,\mathbf{p}}=V$ and $V_{b,\mathbf{p}}=0$, for adsorption on
top of an $A$ site and $V_{a,\mathbf{p}}=0$ and $V_{b,\mathbf{p}}=V$
for a $B$ site. When the adatom sits at the center of the honeycomb
hexagons $(H$-site), the strengths of hybridization with the six
nearest carbon atoms in the tight-binding description depend explicitly
on the symmetry of the orbital: for example, for $s$-wave orbitals,
$V_{x,i}\equiv V$ by symmetry, whereas for in-plane $f$-wave orbitals,
the hybridization amplitudes are anti-symmetric on the two sublattices,
$V_{a,i}=-V_{b,i}=V$\textit{\emph{. In the first case ($s$-wave),
$V_{a,\mathbf{p}}=V\phi_{\mathbf{p}}^{*}$ and $V_{b,\mathbf{p}}=V\phi_{\mathbf{p}}$
whereas in the second ($f$-wave) $V_{a,\mathbf{p}}=V\phi_{\mathbf{p}}^{*}$
and $V_{b,\mathbf{p}}=-V\phi_{\mathbf{p}}$. In the case of a $d_{x^{2}-y^{2}}$
orbital $V_{x,1}=V$, $V_{x,2}=V_{x,3}=-V/2$, whereas for a $d_{xy}$
orbital, $V_{x,1}=0$ and $V_{x,2}=-V_{x,3}=V$, and so on, as illustrated
in Fig. \ref{fig:Illustration-of-d-wave}. Other interesting cases
include for instance substitutional impurities ($S$-sites)\cite{kras},
where $V_{a,i}=0$ for adatoms sitting on $A$ sites and $V_{b.i}=0$
for substitutional adatoms on $B$ sites (see Fig. \ref{fig:C_{3v}-invariant-orbitals}).
A similar description can be for instance applied for adatoms sitting
on bond sites in between two neighbor carbon atoms. }}

Finally, the last term in Eq. (\ref{eq:H1}) includes the Coulomb
energy $(U)$ and exchange energy $(J)$ for the electrons in the
different orbitals\cite{Coqblin}, \begin{eqnarray}
H_{U} & = & \frac{1}{2}\sum_{\sigma}\sum_{m^{\prime},m\neq m^{\prime}}(U_{mm^{\prime}}-J_{mm^{\prime}})f_{m,\sigma}^{\dagger}f_{m,\sigma}f_{m^{\prime},\sigma}^{\dagger}f_{m^{\prime},\sigma}\nonumber \\
 &  & +\sum_{m,m^{\prime}}U_{mm^{\prime}}f_{m,\uparrow}^{\dagger}f_{m,\uparrow}f_{m^{\prime},\downarrow}^{\dagger}f_{m^{\prime},\downarrow},\label{eq:Hu3}\end{eqnarray}
which can be decomposed at the mean-field level into $H_{U}=\sum_{m,\sigma}f_{m,\sigma}^{\dagger}f_{m,\sigma}[\sum_{m^{\prime}}U_{mm^{\prime}}n_{m^{\prime},-\sigma}+\sum_{m^{\prime}\neq m}(U_{mm^{\prime}}-J_{mm^{\prime}})n_{m,^{\prime}\sigma}]$,
where \begin{equation}
n_{m,\sigma}=\langle f_{m,\sigma}^{\dagger}f_{m,\sigma}\rangle\label{eq:n}\end{equation}
 is the occupation of the orbital with angular momentum projection
$m$, and spin $\sigma$. The summation is carried over the degenerate
orbitals in a given irredubile representation, which in graphene correspond
to the doublets $m=\pm|l_{z}|$, with $|l_{z}|=0,1,\ldots l$. The
mean field interaction can be absorbed into the definition of the
localized energy level in Eq. (\ref{Hf}), \begin{equation}
H_{f}=\sum_{\sigma,m}\epsilon_{m,\sigma}\, f_{m,\sigma}^{\dagger}f_{m,\sigma}\,,\label{eq:Hu2}\end{equation}
 where\[
\epsilon_{m,\sigma}\equiv\epsilon_{0}+\sum_{m^{\prime}}U_{mm^{\prime}}n_{m^{\prime},-\sigma}+\sum_{m^{\prime}\neq m}(U_{mm^{\prime}}-J_{mm^{\prime}})n_{m,^{\prime}\sigma}\]
 is the spin dependent renormalized energy of the localized states
in a given irreducible representation.

\section{Local magnetic moments}

The formation of local magnetic moments can be addressed by the self
consistent calculation of the occupation for up and down spin states
in the different orbitals, which follows from integrating the DOS
from the bottom of the band up to the Fermi level $\mu$,\cite{anderson61,uchoa}
\begin{equation}
n_{m,\sigma}=-\frac{1}{\pi}\int_{-\infty}^{\mu}\mbox{d}\omega\,\mbox{Im}G_{ff,m,\sigma}^{R}(\omega)\label{eq:n1}\end{equation}
 where\begin{equation}
G_{ff,m,\sigma}^{R}(\omega)=\left[\omega-\epsilon_{m,\sigma}-\Sigma_{ff,m}(\omega)+i0^{+}\right]^{-1}\label{Gff2}\end{equation}
 is the retarded Green's function of the localized electrons, $G_{ff,\sigma}(\tau)=-\langle T[f(\tau)f^{\dagger}(0)\rangle$,
and \begin{equation}
\Sigma_{ff,m}(\omega)=\sum_{x,y}\sum_{\mathbf{p}}\left[V_{x,\mathbf{p}}^{(m)}\right]^{*}G_{xy}^{0\, R}(\mathbf{p},\omega)V_{y,\mathbf{p}}^{(m)}\label{SEH0}\end{equation}
 is the self-energy of the $f$-electrons, with $x=a,b$. $\hat{G}_{x,y}^{0\, R}$
are the matrix elements of the retarded Green's function of the itinerant
electrons in graphene, $G_{aa,\mathbf{p}}^{0}(\tau)=-\langle T[a_{\mathbf{p}}(\tau)a_{\mathbf{p}}^{\dagger}(0)]\rangle$
and so on, which are are defined by \begin{equation}
\hat{G}(\mathbf{p},i\omega)=\frac{1}{i\omega-\hat{H}}\,,\label{eq:G1}\end{equation}
 where \begin{equation}
\hat{H}=-t\left(\begin{array}{cc}
0 & \phi_{\mathbf{p}}\\
\phi_{\mathbf{p}}^{*} & 0\end{array}\right),\label{eq:H}\end{equation}
 is the tight-binding Hamiltonian matrix. More explicitly,\begin{eqnarray}
G_{xy}^{0,R}(\mathbf{p},\omega) & = & \frac{1}{2}\sum_{\alpha=\pm}\frac{\mathbf{1}+\alpha\hat{\sigma}_{xy,\mathbf{p}}}{\omega-t\alpha|\phi_{\mathbf{p}}|+i0^{+}}\label{Gaa0}\end{eqnarray}
 where \begin{equation}
\hat{\sigma}_{xy,\mathbf{p}}\equiv\frac{\mbox{Re}(\phi_{\mathbf{p}})\sigma_{xy}^{1}-\mbox{Im}(\phi_{\mathbf{p}})\sigma_{xy}^{2}}{|\phi_{\mathbf{p}}|}\,,\label{eq:sigma3}\end{equation}
 $\mathbf{1}$ is the identity matrix and $\sigma^{1}$ and $\sigma^{2}$
are off-diagonal $2\times2$ Pauli matrices, namely $\sigma_{ab}^{1}=\sigma_{ba}^{1}=1$
and $\sigma_{ab}^{2}=-\sigma_{ba}^{2}=-i$.

\begin{figure}
\begin{centering}
\includegraphics[scale=0.2]{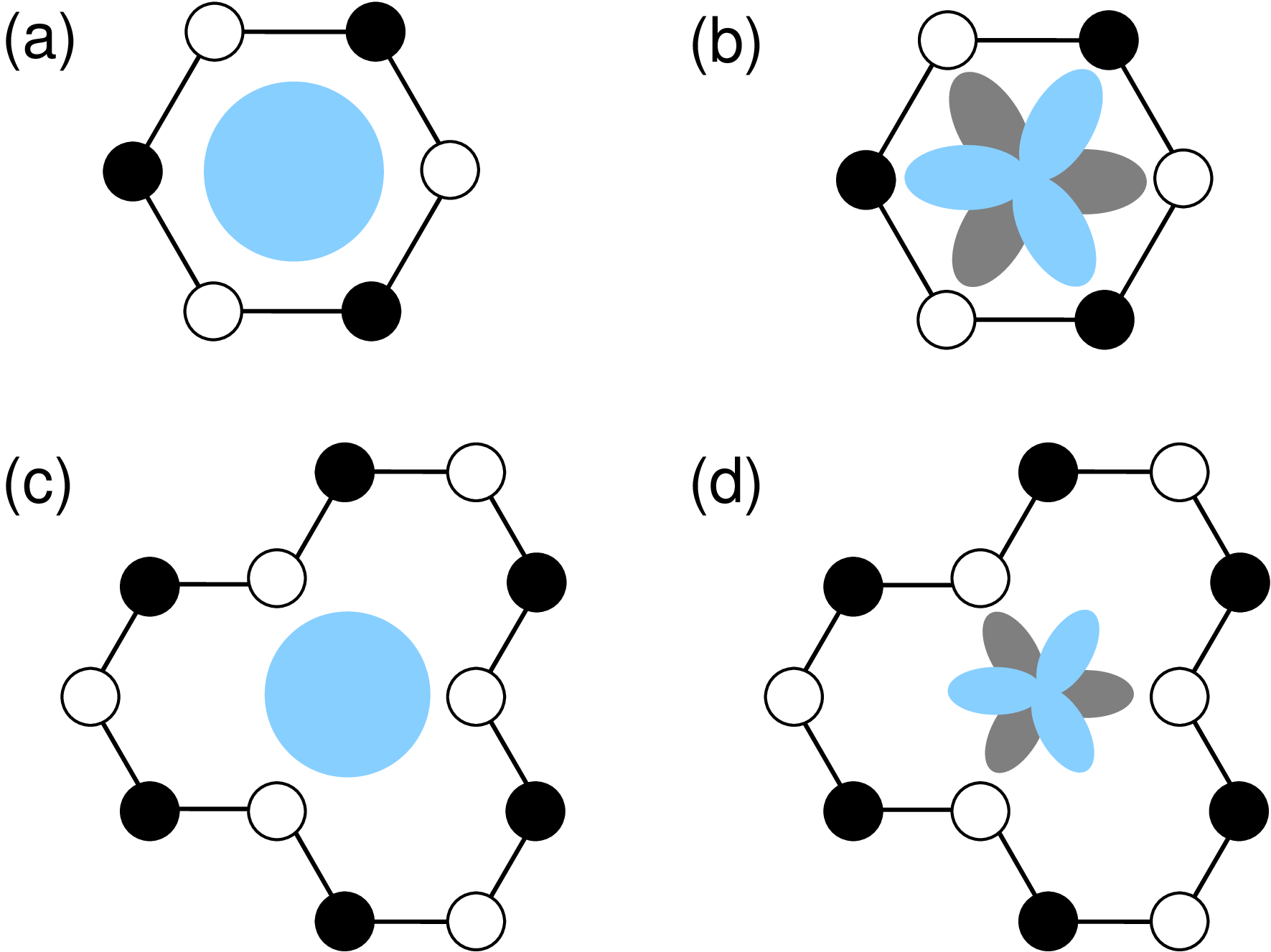}
\par\end{centering}

\caption{{\small $C_{3v}$ invariant orbitals for adatoms sitting in the center
of the hexagon, on $H$ sites (top), and in substitutional ($S$)
sites (bottom). On the left: $s$-wave orbitals, with zero angular
momentum ($m=0$); on the right: in-plane $f$-wave orbitals ($m=\pm3$).
In the two cases, the adatoms hybridize equally with the carbon atoms
on the same sublattice.\label{fig:C_{3v}-invariant-orbitals}}}
\end{figure}

The Green's function of the localized electrons can be written more
explicitly in the following form:\begin{equation}
G_{ff,m,\sigma}^{R}(\omega)=\frac{1}{\omega Z_{m}^{-1}(\omega)-\epsilon_{m,\sigma}+i\Delta^{(m)}(\omega)+i0^{+}}\,,\label{eq:Gff3}\end{equation}
 where \begin{equation}
Z_{m}^{-1}(\omega)=1-\frac{1}{\omega}\mbox{Re}\Sigma_{ff,m}(\omega)\label{eq:Z}\end{equation}
 gives the quasiparticle residue and $\Delta(\omega)\equiv-\mbox{Im}\Sigma_{ff}(\omega)$
is the level broadening of the localized state\cite{Uchoa3},\begin{eqnarray}
\Delta^{(m)}(\omega) & = & \left[V^{(m)}\right]^{2}\sum_{\mathbf{p},\alpha}|\Theta_{\alpha,\mathbf{p}}^{(m)}|^{2}\delta(\omega-\alpha|\phi_{\mathbf{p}}|)\,,\label{eq:Delta}\end{eqnarray}
 which is defined in terms of the generic tight-binding phases,\begin{equation}
\Theta_{\alpha,\mathbf{p}}^{(m)}\equiv\frac{1}{\sqrt{2}V^{(m)}}\left(V_{b,\mathbf{p}}^{(m)}+\alpha\frac{\phi_{\mathbf{p}}^{*}}{|\phi_{\mathbf{p}}|}V_{a,\mathbf{p}}^{(m)}\right)\,,\label{eq:Theta}\end{equation}
 where $V^{(m)}\equiv\mbox{max}(V_{x,i}^{(m)})$.

Those phases depend explicitly on the symmetry of the localized orbital,
which reflect in the relative amplitudes of hybridization with the
surrounding carbon atoms, and also on the relative position of the
adatom with respect to the sublattices, i.e, if the adatom sits on
top of a carbon, in the center of the honeycomb hexagon, on a bridge
site or else in a substitutional site. This formulation is completely
general and can be easily extended to include for instance substitutional
impurities in double vacancies.

In the scenario where the adatom sits on top of a carbon atom, the
level broadening is given by $\Delta(\omega)=\pi V^{2}\rho(\omega)$,
where $\rho(\omega)=|\omega|/D^{2}$ is the DOS in graphene in the
linear portion of the spectrum, and therefore $\Delta(\omega)$ scales
linearly with energy. For adatoms that sit on $H$ or $S$ sites,
the scaling analysis of the level broadening allows a classification
in two symmetry groups, depending on either if the $C_{3v}$ point
group symmetry of the honeycomb sublattice is preserved by the adatom
or not, as previously mentioned in the introduction. As illustrated
in Fig. \ref{fig:C_{3v}-invariant-orbitals}, when the electrons hop
in and out of an adatom sitting on $H$ or $S$ sites, they collect
phases which give rise to quantum mechanical interference among the
possible hybridization paths. When the amplitudes of hybridization
of a localized orbital with the three surrounding carbons on the same
sublattice are identical, in which case the $C_{3v}$ point group
symmetry of sublattice $x$ is preserved, the hopping phases interfere
and give rise to an anomalous energy scaling of the hybridization,
whose modulus scales now in the same way as the Kinetic energy, $|V_{x,\mathbf{p}}|\propto|\phi_{\mathbf{p}}|$.
In that case, the level broadening scales as\cite{Uchoa2,Uchoa3}
\begin{equation}
\Delta^{(m)}(\omega)\approx\pi[V^{(m)}]^{2}\rho(\omega)\frac{|\omega|^{2}}{t^{2}}\label{eq:Delta2}\end{equation}
 at low energy, as opposite to the conventional case where this interference
is frustrated and $|V_{x,\mathbf{p}}|$ scales to a constant near
the Dirac points. In the later, $\Delta(\omega)\propto\rho(\omega)$
corresponds to the standard case, whereas in the former case the damping
is super-linear. The first class of orbitals, which we will refer
as type II orbitals, include $m=0$ and $m=3$ angular momentum states,
such as in $s$ and in-plane $f$-wave orbitals. The standard {}``ohmic''
class (type I) by its turn is described by adatoms on top carbon sites
and $m=\pm1$ and $m=\pm2$ angular momentum orbitals on $H$ or $S$
sites. To be more concrete, the class of type I orbitals is represented
by adatoms that sit on top of a carbon atom, in which case the orbital
symmetry is not particularly important, and also by adatoms siting
at $H$ or $S$ sites with localized orbitals in the $E_{1}$($d_{xz},$$d_{yz}$)
(i.e $m=\pm1)$ and $E_{2}$$(d_{xy},d_{x^{2}-y^{2}})$ ($m=\pm2)$
representations of $d$-wave orbitals and also $f_{xz^{2}}$, $f_{yz^{2}}$,
$f_{xyz}$, $f_{z(x^{2}-y^{2})}$ orbitals in $H$/$S$ sites. The
class of type II orbitals is described by $s$, $d_{zz}$, $f_{z^{3}}$
orbitals, where $m=0$, and also by $f_{x(x^{2}-3y^{2})}$, and $f_{y(3x^{2}-y^{2})}$
orbitals ($m=\pm3)$ in $H$ or $S$ sites. The anomalous scaling
of the level broadening in Eq. (\ref{eq:Delta2}) has been verified
explicitly by ab initio methods, in particular for the $d_{zz}$ orbital
of Co on graphene\cite{Wehling2}.

The self-energy of the localized electrons {[}see Eq. \ref{SEH0}{]}
can be more explicitly written in the form \begin{equation}
\Sigma_{ff,m}(\omega)=-\omega\left[Z_{m}^{-1}(\omega)-1\right]-i\Delta^{(m)}(\omega)|\theta(D-|\omega|)\,.\label{eq:SelfI}\end{equation}
 The density of states of the localized level, $\rho_{ff,,m,\sigma}(\omega)=-1/\pi\mbox{Im}\, G_{ff,m,\sigma}^{R}(\omega)$
follows from the substitution of Eq. (\ref{eq:SelfI}) into Eq. (\ref{Gff2}),
\begin{equation}
\rho_{ff,m,\sigma}(\omega)=\frac{1}{\pi}\frac{\Delta^{(m)}(\omega)\theta(D-\vert\omega\vert)}{[\omega Z_{m}^{-1}(\omega)\ -\epsilon_{m,\sigma}]^{2}+\left[\Delta^{(m)}(\omega)\right]^{2}}\,.\label{rhoff}\end{equation}

In the linear cone approximation, where the spectrum is linearized
around the Dirac points, $t|\phi_{\mathbf{K}+\mathbf{p}}|\to vp$
up to the cut-off of the band $D$, with $v\approx6\mbox{eV\AA}$
as the Fermi velocity, the level broadening for orbitals of type I
is \begin{equation}
\Delta_{I}(\omega)\equiv\Delta_{0}|\omega|\label{eq:Delta0}\end{equation}
 at low energies, where $\Delta_{0}=\pi(V/D)^{2}$ is the dimensionless
hybridization parameter, and \begin{equation}
Z_{I,m}^{-1}(\omega)=(\Delta_{0}^{(m)}/\pi)\ln\left|1-D^{2}/\omega^{2}\right|,\label{eq:Z1}\end{equation}
 implying that the quasiparticle residue $Z\to0$ vanishes logarithmically
at low energy.

\begin{figure}[t]
\begin{centering}
\includegraphics[scale=0.37]{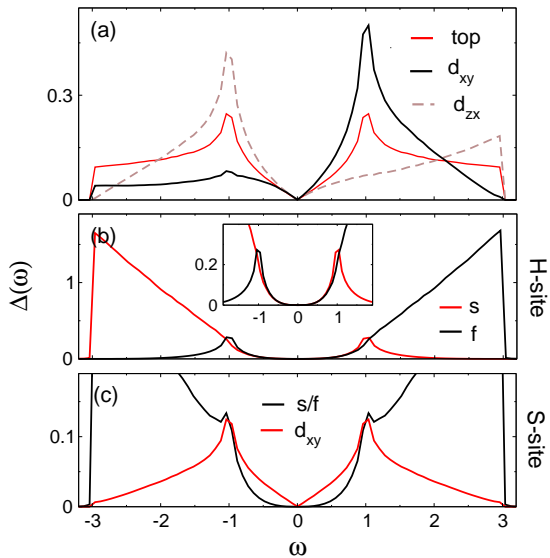} 
\par\end{centering}

\caption{{\small Level broadening $\Delta(\omega)$ as a function of energy,
$\omega$, for different orbital symmetries. All energies in units
of the hopping energy $t$ ($V/t=1/3$). a) Type I orbitals for adatoms
sitting on a top carbon site (light/red solid curve) and for $d_{xy}$-wave
(black line) and $d_{zx}$-wave orbitals (dashed line) on $H$ sites.
b) Type II orbitals on $H$ sites. Black curve: in-plane $f$-wave
orbital; light/red curve: $s$-wave orbital. Inset: low energy scaling
of the level broadening, $\Delta(\omega)\propto|\omega|^{3}$ for
$|\omega|/t<1$ (see text). c) Substititutional $s/f$-wave orbitals
(black curve) and $d_{xy}$-wave orbital (light/red curve), on $S$
sites. \label{fig:Level-broadening-Delta}}}
\end{figure}

In the case of super-linear damping, for type II orbitals, the level
broadening scales with the \emph{cube} of the energy within the linear
cone approximation, \begin{equation}
\Delta_{II}(\omega)=N_{s}\Delta_{0}|\omega|^{3}/t^{2}\,,\label{eq:Delta3}\end{equation}
 and can be orders of magnitude smaller than in the linear case when
$|\epsilon_{0}|\ll t$. $N_{s}=1,2$ correspond to the number of sublattices
the adatom effectively hybridizes ($N_{s}=2$ for $H$ sites and $N_{s}=1$
for $S$ sites, as shown in Fig. \ref{fig:C_{3v}-invariant-orbitals}).The
quasiparticle residue, $Z_{\sigma}$, in this approximation is given
by\begin{equation}
Z_{II,m}^{-1}(\omega)=1+\frac{N_{s}\Delta_{0}^{(m)}}{\pi t^{2}}\left[D^{2}+\omega^{2}\ln|1-D^{2}/\omega^{2}|\right].\label{eq:Z3}\end{equation}
In all cases, the level broadening \textcolor{black}{can be severely
affected in the presence of out-of-plane magnetic fields, which produce
Landau level quantization and non-linear orbital magnetization effects
in graphene \cite{Betouras}.}

In Fig. \ref{fig:Level-broadening-Delta}, we show the energy scaling
of the level broadening for the different orbital symmetries. Particle-hole
symmetry is preserved for adatoms on top carbon sites, where $\Delta(\omega)$
follows the DOS, and also for adatoms on $S$ sites, which effectively
hybridize with only one sublattice. For adatoms on $H$ sites, which
hybridize with the two sublattices, particle-hole symmetry is explicitly
broken in the high energy sector ($|\omega|\gtrsim t$) by the off
diagonal matrix elements of the hybridization (at low energy, the
off diagonal terms average to zero in the momentum integrals). In
particular, $d_{x^{2}-y^{2}}$, $d_{xy}$-wave orbitals (where $V_{a,i}=V_{b,i}$)
are strongly damped when the energy of the localized state is far
above the Dirac point ($\omega>t$), but otherwise are weakly damped
at negative energy states (black solid curve in Fig. \ref{fig:Level-broadening-Delta}a).
Conversely, $d_{zx}$ and $d_{zy}$-wave orbitals (where $V_{a,i}=-V_{b,i}$)
show the opposite trend, in agreement with \emph{ab initio} calculations
for Co adatoms in graphene\cite{Wehling2,Jacob}. In the same way,
$s$ and in-plane $f$-wave orbitals, which couple symmetrically and
anti-symmetrically with the two sublattices respectively, show a strong
particle-hole asymmetry at high energies, as depicted in Fig. \ref{fig:Level-broadening-Delta}b.
In Fig.\ref{fig:Level-broadening-Delta}c, we show the level broadening
for the substitutional case, where particle-hole symmetry is restored.
In all cases, the peaks at $|\omega|=t$ are divergences which are
reminiscent of the logarithmic singularity of the DOS around the $M$
point of the BZ.

\subsection{\textcolor{black}{Single orbital picture: The case of spin $1/2$
adatoms}}

\textcolor{black}{The arguments outlined so far are completely general,
and apply to any adatom in graphene containing localized electronic
states. Now, for convenience and simplicity, we will restrict our
analysis to adatoms that can be described by an effective single orbital
picture.}

\begin{figure}[t]
\begin{centering}
\includegraphics[scale=0.4]{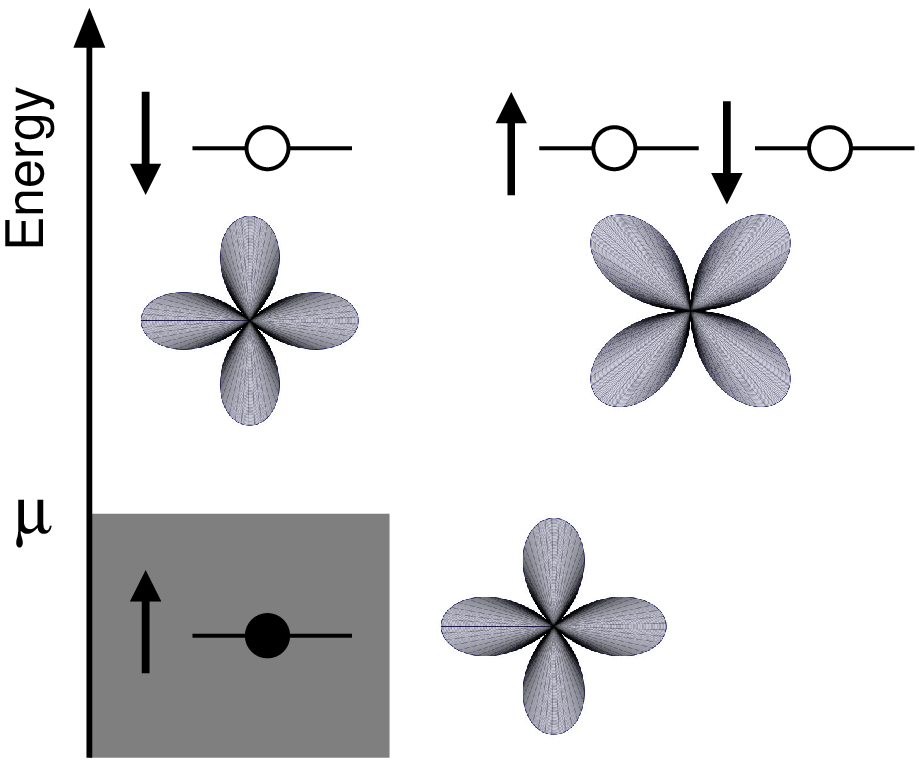} 
\par\end{centering}

\caption{\textcolor{black}{\small Maximal spin and orbital polarization for
degenerate doublet states, say $d_{x^{2}-y^{2}}$ and $d_{xy}$ $(m=\pm2)$.
When the bare levels are occupied $\mu-\epsilon_{0}>0$ and the virtual
doubly occupied state is empty, $\epsilon_{0}+U\gg\mu$, where $U$
is the Coulomb energy, the energy of the doublet state is maximally
polarized in the regime where $U\gg J,\,\Delta$, where $J$ is the
exchange energy and $\Delta$ is the level broadening due to the hybridization
with the bath. In this limit, the spin and orbital polarized levels
have occupation $n_{m,\sigma}\sim1$, and $n_{m,-\sigma}=$ $n_{-m,\sigma}=n_{-m,-\sigma}\sim0$
(spin $1/2$) (see Ref. 27) with $\sigma=\uparrow\downarrow$. \label{fig:Maximal spin and orbit}}}
\end{figure}

\textcolor{black}{Let us consider for instance two degenerate orbitals
contained in a given irreducible representation of the honeycomb lattice,
say $d_{xy}$ and $d_{x^{2}-y^{2}}$, with $m=\pm2$. We assume that
the bare degenerate energy levels are occupied, $\epsilon_{0}-\mu<0$,
and the virtual doubly occupied states are empty, $\epsilon_{0}+U\gg\mu$.
In the situation where Coulomb repulsion in the orbitals $U$ is much
larger that the exchange coupling $J$ and the level broadening $\Delta$
due to the hybridization of the orbitals, namely $U\gg J,\,\Delta$,
the lowest energy solution for a doublet is a maximally polarized
state where one orbital is fully spin polarized (say $d_{x^{2}-y^{2}}$)
and the other is empty (hence with total spin $1/2$ and total charge
of one electron)\cite{Coqblin}, as illustrated in Fig. \ref{fig:Maximal spin and orbit}.
In this regime, where only one orbital of the doublet is occupied,
the energy separation between the orbitals is set by the Coulomb interaction,
$U$, which is typically of the order of a few eV. The spin and the
charge of the polarized orbitals will fluctuate among four minima,
which describe the four possible degenerate configurations with the
lowest energy, namely \begin{equation}
n_{m,\sigma}\sim1,\qquad n_{m,-\sigma}=n_{-m,\sigma}=n_{-m,-\sigma}\sim0,\label{eq:n-1}\end{equation}
for $m=\pm|m|$ and $\sigma=\uparrow,\downarrow$. The orbital degeneracy
of the four minima can be lifted through a Jahn -Teller effect distortion
created by local lattice deformations. }

\textcolor{black}{In real crystals, the adatoms are known to locally
deform the lattice, and those deformations extend over several lattice
sites\cite{Liu2}. In the situation where the adatoms are randomly
distributed, those distortions will displace the adatoms from the
high symmetry positions and create crystalline field anisotropies.
Those anisotropies will select one of the two degenerate orbitals
and {}``freeze'' their occupation at energy scales smaller that
the crystalline field anisotropy energy. When this criterion is fulfilled
and $U\gg J,\,\Delta$, the electronic transitions between the orbitals
are suppressed and the effective Hamiltonian of the spin polarized
orbital can be approximated for all purposes by the Hamiltonian of
the}\textcolor{black}{\emph{ single}}\textcolor{black}{{} orbital problem,
up to small corrections due to the direct Coulomb interaction between
the orbitals. We will assume the local deformations in graphene to
be small but enough to lift the orbital degeneracies of the adatoms
beyond the experimental STM energy resolution.}

The single orbital problem was described in the original work by Anderson\cite{anderson61}.
The Coulomb Hamiltonian of the spin polarized orbital is approximately
described by \begin{eqnarray}
H_{U} & = & U\, f_{\uparrow}^{\dagger}f_{\uparrow}f_{\downarrow}^{\dagger}f_{\downarrow},\label{eq:Hu2-1}\end{eqnarray}
and orbital indexes $(m)$ are suppressed everywhere else in Hamiltonian
(\ref{eq:H1}). The energy of the localized orbital becomes \begin{equation}
H_{f}=\sum_{\sigma}\epsilon_{\sigma}\, f_{\sigma}^{\dagger}f_{\sigma}\,,\label{eq:Hf3}\end{equation}
where $\epsilon_{\sigma}=\epsilon_{0}+Un_{-\sigma}$ gives the energy
of the spin polarized level. 

\begin{figure}[b]
\begin{centering}
\includegraphics[scale=0.33]{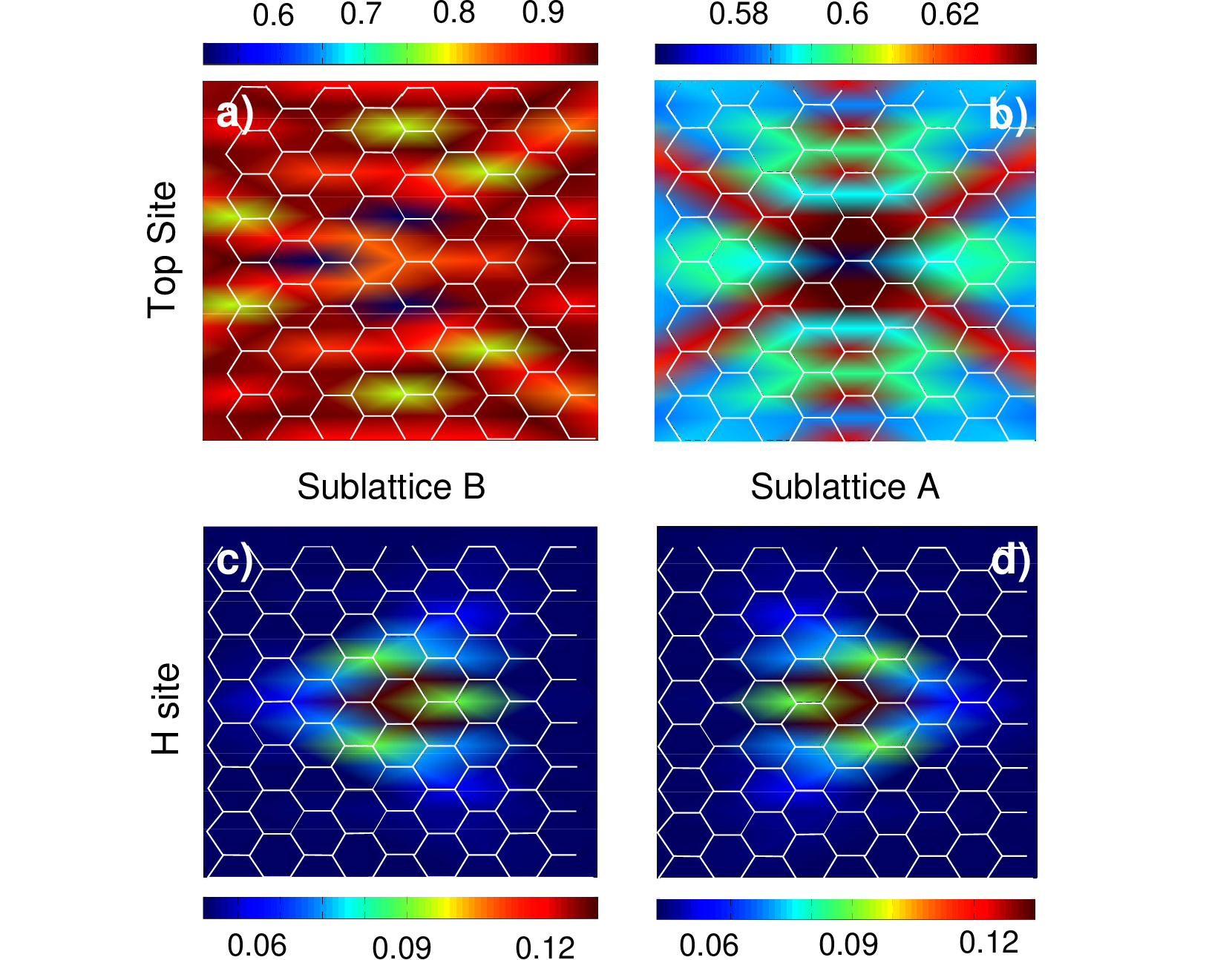} 
\par\end{centering}

\caption{{\small Comparison of the LDOS between the two different sublattices.
Top panels: Energy integrated LDOS around an adatom (center) sitting
on a top carbon adatom site. Scans for a) the opposite sublattice
of the impurity and b) for the same sublattice. Lower panels: Energy
integrated LDOS around a localized orbital (center) with }\emph{\small s}{\small -wave
symmetry, when the adatom sits in the center of a honeycomb hexagon
($H$ site). c) scans for sublattice $A$ and d) $B$. The two scans
are related by a $\pi$-rotation. \label{fig:Integrated-LDOS-around}}}
\end{figure}

\begin{figure*}
\begin{centering}
\includegraphics[scale=0.33]{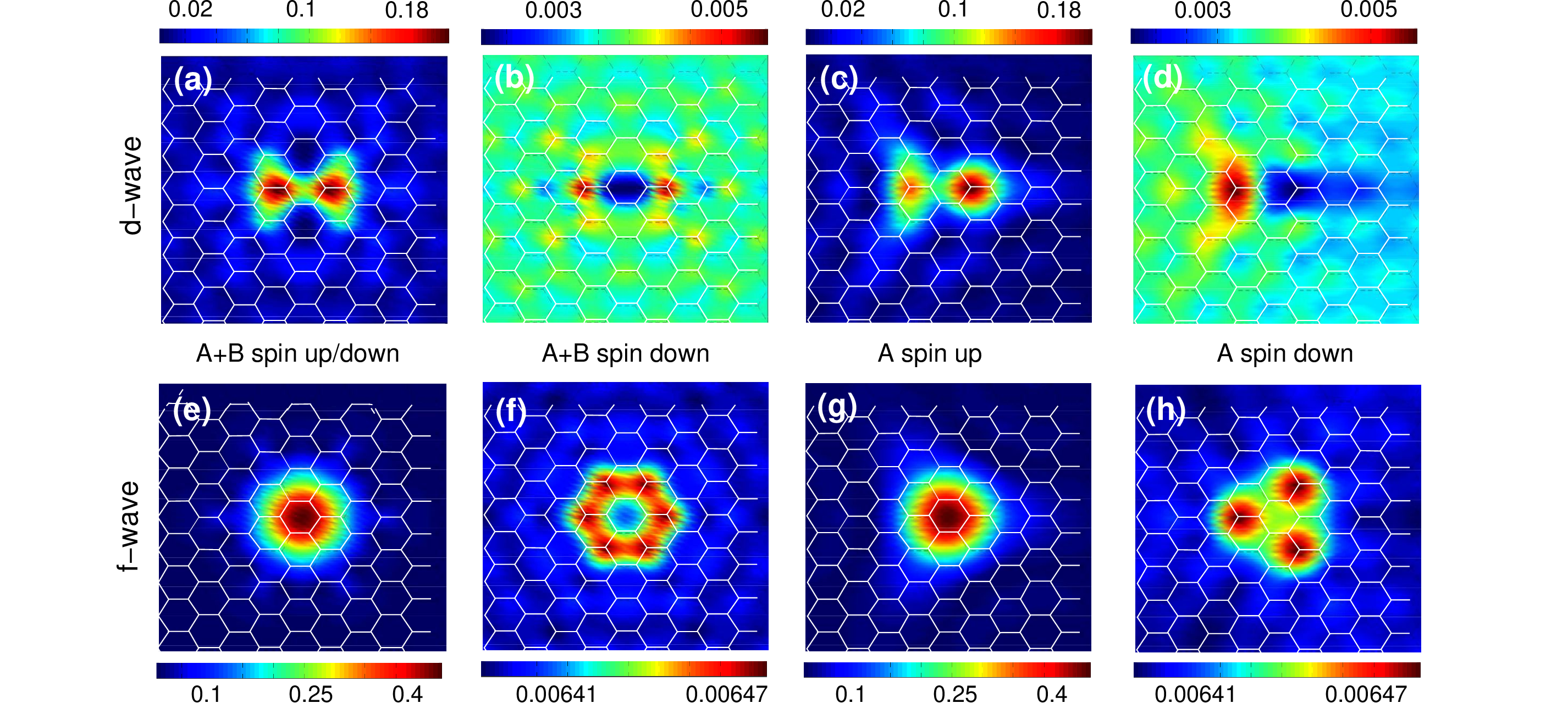} 
\par\end{centering}

\caption{{\small LDOS around the adatom (center) at a fixed energy ($\omega=-0.2$eV)
for adatoms on $H$ sites. The top row corresponds to a $d_{x^{2}-y^{2}}$
orbital and the bottom row to an in-plane $f$-wave orbital. a) and
e): Total LDOS; b) and f): LDOS for the minority spin; c) and g) LDOS
for the majority spin on sublattice $A$; d) and h) LDOS for the minority
spin on sublattice $A$.\label{fig:LDOS-around-the}}}
\end{figure*}

In the single orbital problem, as in the degenerate case, the occupation
for up and down spin states can be self-consistently calculated at
the mean field level from Eq. (\ref{eq:n1}). The emergence of a local
magnetic moment follows from the appearance of a spin polarized state
below the Fermi level, say, at energy $\epsilon_{0}+n_{\uparrow}U$,
and a virtual (empty) state at $\epsilon_{0}+n_{\downarrow}U$ for
the majority spin, with $n_{\uparrow}+n_{\downarrow}\leq1$ due to
the Pauli principle. 

The analysis about the formation of local magnetic moments and the
zero dimensional phase diagram that comes out of the single orbital
picture has been discussed in detail in ref. \onlinecite{uchoa}
for the case of type I orbitals. For type II orbitals, the physics
is qualitatively similar, except for the fact that the formation of
a local magnetic moment becomes exceedingly easy, even at small $U$,
due to the fact that the broadening of the level can be negligibly
small when $\epsilon_{0}/t\ll1$. 

From now on, we will drop the orbital indexes $m$ and consider only
spin polarization effects on a given orbital.

\section{Local DOS}

The local DOS around the impurity can be computed directly from the
diagonal matrix elements of the electronic Green's function in graphene
in the presence of the adatom,\begin{equation}
\rho_{x}(\mathbf{r},\omega)=-\frac{1}{\pi}\mbox{Im}\sum_{\sigma}\sum_{\mathbf{p},\mathbf{p}^{\prime}}G_{xx,\sigma}^{R}(\mathbf{p},\mathbf{p}^{\prime},\omega)\mbox{e}^{i(\mathbf{p}-\mathbf{p}^{\prime})\cdot\mathbf{R}}\,,\label{eq:rho0}\end{equation}
 where \begin{equation}
G_{xy,\sigma}(\mathbf{p},\mathbf{p}^{\prime},i\omega)=\delta_{\mathbf{p},\mathbf{p}^{\prime}}G_{xy}^{0}(p)+\Lambda_{x}(p)G_{ff,\sigma}(i\omega)\bar{\Lambda}_{y}(p^{\prime})\,,\label{eq:Gxx01}\end{equation}
 and \begin{eqnarray}
\Lambda_{x}(p) & \equiv & \sum_{y=a,b}G_{xy}^{0}(p)V_{y,\mathbf{p}}\label{eq:Lambda2}\\
\bar{\Lambda}_{x}(p) & \equiv & \sum_{y=a,b}V_{y,\mathbf{p}}^{*}G_{yx}^{0}(p)\,,\label{Lambdabar}\end{eqnarray}
 with $V_{a,\mathbf{p}}$ and $V_{b,\mathbf{p}}$ defined in Eq. (\ref{eq:Vb})
and (\ref{eq:Va}). 

In Fig. \ref{fig:Integrated-LDOS-around} we show the topography maps
around the impurity, which describe the local DOS integrated in energy.
We use the set of parameters $V=1$eV, $U=1$eV, $\mu=0.1$eV and
$\epsilon_{0}=-0.5$eV. Those energy scales describe the typical order
of magnitude for the charging energy $U$ and the hybridization $V$
of transition metals in graphene \cite{Wehling}. In the case where
the adatoms sit on top of a carbon site on a given sublattice (top
panels), the maps show a clear asymmetry between the integrated LDOS
of the two different sublattices. The pattern on the opposite sublattice
of the impurity (Fig. \ref{fig:Integrated-LDOS-around}a) has a lower
point group symmetry than in the same sublattice (Fig. \ref{fig:Integrated-LDOS-around}b),
what comes from the fact that the adatom in this case has only three
nearest neighbor carbon sites but six next-nearest neighbor ones.
For adatoms sitting in the center of the honeycomb hexagon (lower
panels), there is no distinction between the patterns of the two different
sublattices, except for a rotation of $\pi$. Fig. \ref{fig:Integrated-LDOS-around}c
and d depict the integrated LDOS for an $s$-wave orbital ($m=0$)
sitting on an $H$ site. The intensity of the integrated LDOS maps
is also much weaker in the lower panels compared to the upper ones,
reflecting the fact that the hybridization for $H$ or $S$ sites
is mediated by hopping, and hence weaker than in the top carbon site
case for the same set of parameters. Only recently STM experiments
observed the topography around isolated Fe and Co adatoms in graphene
\cite{Elbo}. X-ray magnetic circular dichroism experiments revealed
that these adatoms have a large local magnetic moment \cite{Elbo}. 

\begin{figure}[t]
\begin{centering}
\includegraphics[scale=0.45]{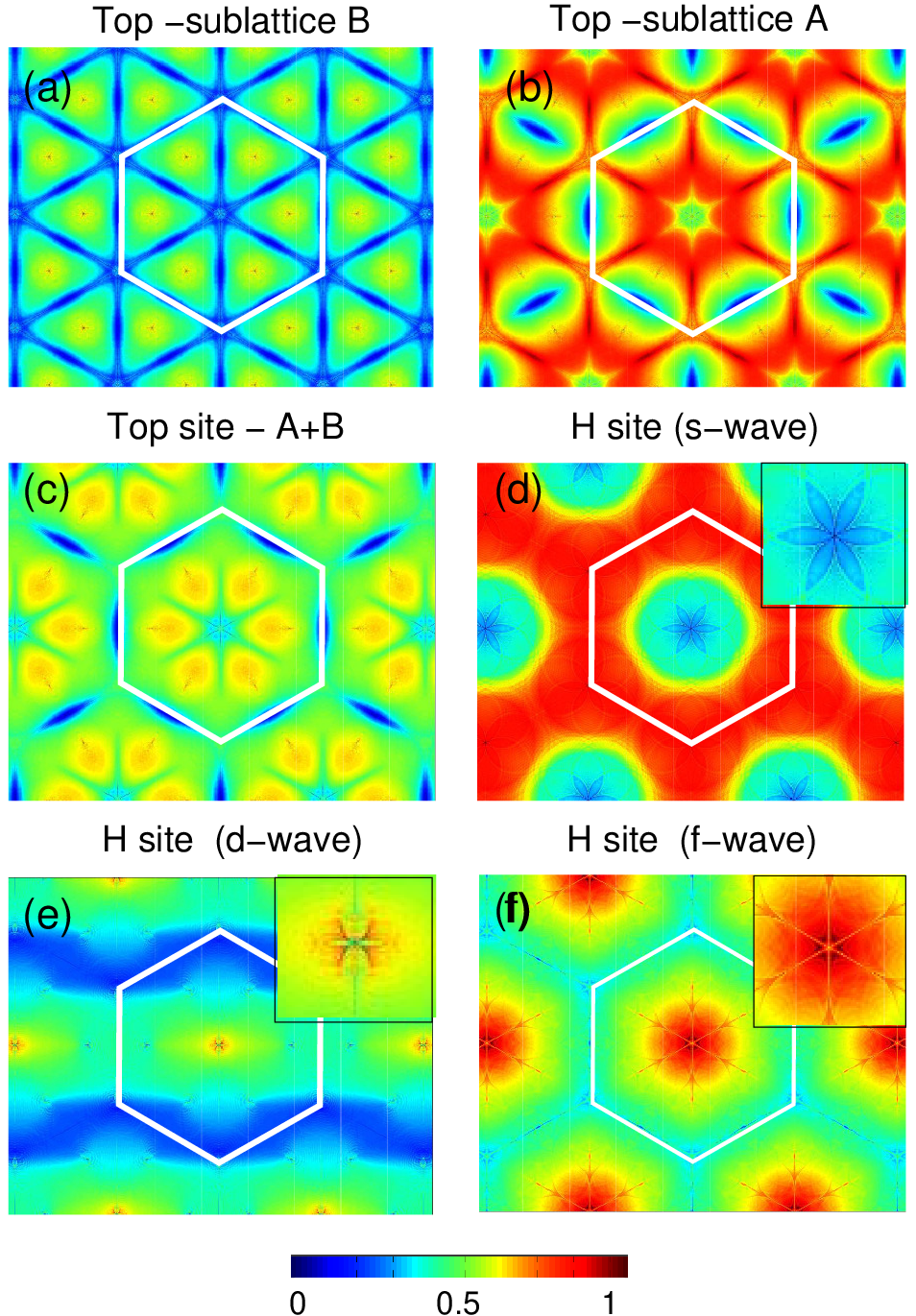} 
\par\end{centering}

\caption{{\small Fourier transform of the energy integrated LDOS around a spin
$1/2$ adatom. Solid hexagon line indicates the Brillouin zone. The
peaks at the center of the zone correspond to forward scattering processes,
whereas the peaks at the corners of the zone ($K$) points correspond
to backscattering between the two valleys. Top panels: adatom on a
top carbon site case; Fourier transform of the LDOS for the a) opposite
and b) same sublattice of the adatom. c) superposition of the patterns
in panels a) and b), for both sublattices. d) $s$-wave orbital at
an $H$ site. Lower panels: e) $d_{x^{2}-y^{2}}$-wave orbital and
f) $f_{x(x^{2}-3y^{2})}$ orbital also at $H$ sites. The insets in
panels d), e) and f) show the details of the forward scattering peaks
at the center of the BZ. \label{fig:FourierTransform}}}
\end{figure}

The analysis of the LDOS also permits to identify the symmetry of
the localized orbital when the adatom sits either on $H$ or $S$
sites. In Fig. \ref{fig:LDOS-around-the} we show the LDOS at fixed
energy for both a $d_{x^{2}-y^{2}}$ orbital (Fig. \ref{fig:LDOS-around-the}a-d)
and a $f_{x(x^{2}-3y^{2})}$ state (Fig. \ref{fig:LDOS-around-the}e-h).
In the former, the orbital $d$-wave symmetry of the localized state
has a clear fingerprint in the induced DOS nearby the adatom. The
signature is specially pronounced when the LDOS is resolved for the
minority spins, as noticed by direct comparison of Fig. \ref{fig:LDOS-around-the}b
and f. In panels c), d) and g), h), we show the distinct patterns
for the LDOS of majority and minority spins on a given sublattice.
In the case of an in-plane $f$-wave state, which explicitly preserves
the point group symmetry of the sublattices, the fingerprint, although
more subtle, can be easily characterized by a Fourier analysis of
the LDOS, which maps the scattering wavevectors responsible for the
emergence of local resonances nearby the adatom.

In Fig. \ref{fig:FourierTransform}, we plot the corresponding maps
of the energy integrated LDOS in the reciprocal space. The solid hexagonal
line indicates the Brillouin zone (BZ). In Fig. \ref{fig:FourierTransform}a,
we show the maps for the opposite sublattice of the impurity, whereas
on Fig. \ref{fig:FourierTransform}b, we depict the Fourier transformed
maps for the same sublattice of the adatom. The central peaks around
the center of the zone ($\Gamma$ point) indicate forward scattering
processes that contribute to the resonant states nearby the adatom,
whereas the peaks centered at the $K$ points at the edges the BZ
indicate backscattering processes, which connect the different valleys.
For the opposite sublattice of the adatom (Fig. \ref{fig:FourierTransform}a),
the backscattering processes at the $K$ point and the forward scattering
ones at $\Gamma$ are significantly attenuated. In the reciprocal
space maps for the \emph{same} sublattice of the impurity (Fig. \ref{fig:FourierTransform}b),
where unitary scattering should dominate, backscattering processes
are strongly enhanced, followed by the presence of subdominant forward
scattering peaks. In panel \ref{fig:FourierTransform}c, we depict
the Fourier transformed map for both sublattices. In those plots (top
carbon case), the amount of scattering at the $M$ points, which indicate
the position of the Van Hove singularities, is weak compared to the
other dominant processes. The shape of the forward scattering peak
at the $\Gamma$ point also reflects the symmetry of the hybridization
matrix elements in the Hamiltonian. In the top carbon case, the $\Gamma$
peak is isotropic.

In Fig. \ref{fig:FourierTransform}d we show the reciprocal space
maps for the energy integrated LDOS for an $s$-wave orbital siting
at an $H$ site. In this case, the height of the central peak is significantly
small compared to the dominant peaks around the $K$ points, indicating
strong enhancement of the backscattering compared to forward scattering
processes. In panel \ref{fig:FourierTransform}e we depict the case
of a $d_{x^{2}-y^{2}}$-wave orbital also at an $H$ site, whereas
in panel \ref{fig:FourierTransform}f we show the signature of an
$f_{x(x^{2}-3y^{2})}$ orbital ($H$ site) in the reciprocal space.
For $d$ and $f$-wave orbitals, destructive interference leads to
attenuation of the backscattering peaks at the $K$ points, in particular
in the $d$-wave case. In the insets of Fig. \ref{fig:FourierTransform}d
and f we show in detail the features of the forward scattering peaks
for $s$-wave and $f_{x(x^{2}-3y^{2})}$-wave orbitals, respectively.
Both peaks reflect the underlying $C_{3v}$ symmetry of the sublattices,
which are incorporated into the hybridization matrix elements of the
Hamiltonian for orbitals of type II. For a $d$-wave orbital (inset
of Fig. \ref{fig:FourierTransform}e), the forward scattering peak
has $C_{2v}$ symmetry.

\section{STM tip effects}

Let us now consider a problem of more practical interest for tunneling
microscopy experiments, where we include an STM tip close to a spin
$1/2$ impurity. The electrons in the metallic tip follow the Hamiltonian
\begin{equation}
H_{t}=\sum_{\mathbf{p}}\epsilon_{\mathbf{p}}c_{\sigma\mathbf{p}}^{\dagger}c_{\sigma\mathbf{p}}\label{Ht}\end{equation}
 where $\epsilon_{\mathbf{p}}=p^{2}/2m$ is the electronic dispersion
of an electron gas, with $m$ the effective mass. The electrons can
tunnel either to the carbon sites in graphene or to the impurity.
In the former case, the tunneling process is described by the Hamiltonian\begin{eqnarray}
H_{g-t} & = & \sum_{\langle i\rangle}\sum_{\sigma}t_{a}a_{\sigma}^{\dagger}(\mathbf{R}_{i})c_{\sigma}(\mathbf{R}_{i}-\mathbf{r})\nonumber \\
 &  & +\sum_{\langle i\rangle}\sum_{\sigma}t_{b}b_{\sigma}^{\dagger}(\mathbf{R}_{i})c_{\sigma}(\mathbf{R}_{i}-\mathbf{r})+h.c.\label{Hg-t}\end{eqnarray}
 where $t_{a},t_{b}$ are the electronic tunneling energy from the
tip to sublattices $A$, and $B$ in graphene, $\langle i\rangle_{t}$
denotes sum over the tip nearest neighbor carbon sites (\textbf{$\mathbf{R}_{i}$})
on a given sublattice, and $\mathbf{r}=(R,z)$ is the position of
the center of the tip, where $\mathbf{R}$ is the horizontal distance
of the tip to the impurity and $z$ is the distance of the tip to
the graphene layer.

The single-particle wave-functions $\psi_{t,\mathbf{p}}(\mathbf{r})$
describing the electronic state at the tip, namely \[
c_{\sigma}(\mathbf{R}_{i}-\mathbf{r})=\sum_{\mathbf{p}}\psi_{t,\mathbf{p}}(\mathbf{R}_{i}-\mathbf{r})c_{\sigma\mathbf{p}}\,,\]
 can be expanded in spherical waves from the center of the tip, $\psi_{t,\mathbf{p}}(r)\propto\mbox{e}^{-\kappa_{\mathbf{p}}r}/r$,
where $c_{\sigma\mathbf{p}}$ is a second quantized operator for the
tip electrons. The factor\cite{Tersoff,plihal} \begin{equation}
\kappa_{\mathbf{p}}=\sqrt{2m(\phi_{t}-\epsilon_{\mathbf{p}})}\label{eq:kappa}\end{equation}
 gives the effective tunneling barrier between the tip and the rest
of the system, and is defined by the electronic work function of the
tip, $\phi_{t}$. Since the single-particle wave functions of the
graphene electrons can be expanded in plane waves, $\psi_{g,\mathbf{k}}(\mathbf{R}_{i})=\mbox{e}^{i\mathbf{k}\cdot\mathbf{R}_{i}}$\cite{Tersoff},
the Hamiltonian (\ref{Hg-t}) becomes \begin{equation}
H_{g-t}=\sum_{\sigma}\sum_{\mathbf{k}\mathbf{p}}\left[t_{a,\mathbf{k}\mathbf{p}}(\mathbf{r})\, a_{\sigma\mathbf{k}}^{\dagger}c_{\sigma\mathbf{p}}+t_{b,\mathbf{k}\mathbf{p}}(\mathbf{r})\, b_{\sigma\mathbf{k}}^{\dagger}c_{\sigma\mathbf{p}}\right]+h.c.\,,\label{Hg-t2}\end{equation}
 where \begin{equation}
t_{x,\mathbf{k}\mathbf{p}}(\mathbf{r})\sim\frac{t_{x}}{z}\mbox{e}^{-\kappa_{\mathbf{p}}z}\mbox{e}^{-i\mathbf{k}\cdot\mathbf{R}}\equiv t_{x,\mathbf{p}}(z)\,\mbox{e}^{-i\mathbf{k}\cdot\mathbf{R}}\,.\label{tr}\end{equation}
 describes the spatially averaged hopping matrix elements between
the tip and graphene, where the position of each of the carbon atoms
underneath the tip is effectively replaced by the in-plane position
of the center of the tip with respect to the impurity.

\begin{figure}[t]
\begin{centering}
\includegraphics[scale=0.26]{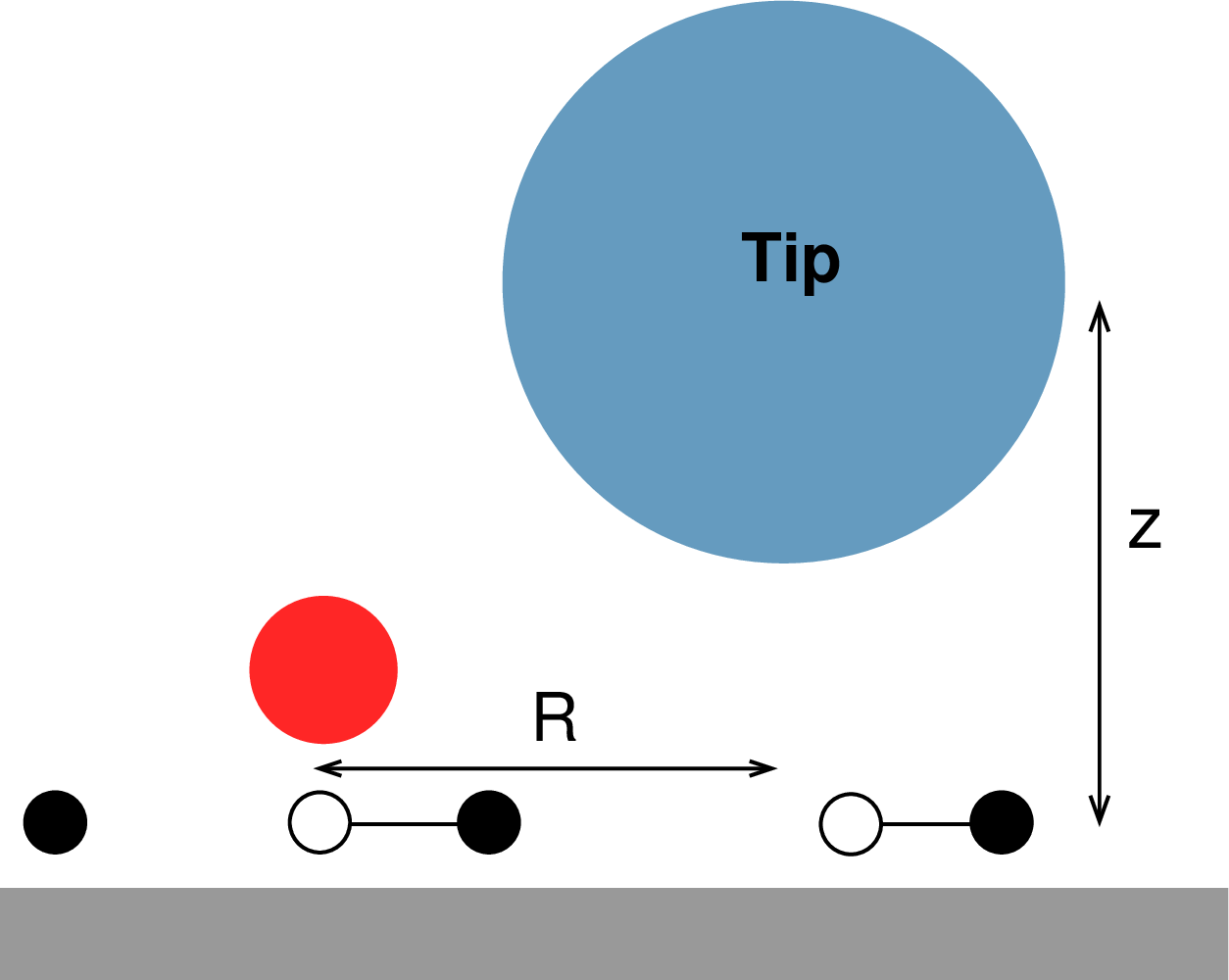} 
\par\end{centering}

\caption{{\small Schematic drawing of the STM tip nearby an adatom (small light
{[}red{]} circle) on top of graphene. Black and white circles: carbon
atoms on sublattices A and B. $R$ is the in-plane distance of the
impurity to the tip and $z$ the out of plane distance from the center
of the tip to the graphene layer. \label{fig:Schematic-drawing-of}}}
\end{figure}

The Hamiltonian for the tunneling from the tip to the impurity is
given by:\begin{equation}
H_{f-t}=t_{f}\sum_{\sigma}f_{\sigma}^{\dagger}c_{\sigma}(\mathbf{r})+h.c.\label{eq:Hf-t}\end{equation}
 where $t_{f-t}$ is the tunneling energy from the tip to the impurity.
In a similar way, we can write \begin{equation}
H_{f-t}=\sum_{\sigma,\mathbf{p}}t_{f,\mathbf{p}}(r)f_{\sigma}^{\dagger}c_{\sigma\mathbf{p}}+h.c.\,,\label{Hf-t}\end{equation}
 where \begin{equation}
t_{f,\mathbf{p}}(r)=t_{f}\,\frac{1}{|\mathbf{r}|}\mbox{e}^{-\kappa_{\mathbf{p}}|\mathbf{r}|}.\label{tf}\end{equation}
 where $|\mathbf{r}|=\sqrt{R^{2}+z^{2}}$ measures the center of the
tip with respect to the position of the impurity. We will assume that
the tip is sufficiently large such that local gating effects due to
the proximity of the tip to graphene can be effectively absorbed into
the local definition of the chemical potential nearby the adatom\cite{Brar2}.

As a brief comment about orders of magnitude for the several quantities,
in most materials, the typical work function $\phi_{t}$ is of the
order of a few eV. In the case where the effective mass $m$ is of
the same order of the bare mass of the electron, $\kappa_{\mathbf{p}}^{-1}$,
translates into a characteristic tunneling length scale topically
larger than 1 nm. Also, since STM tips have a typical radius of the
order of 1nm, $z$, which accounts for the distance between the center
of the tip to graphene is typically a number of the same order. A
more detailed description about the characterization of the tip and
the surface in the STM problem can be found in Ref. \onlinecite{Tersoff}.

\subsection{Green's functions}

Now we generalize the original Hamiltonian of the Anderson problem
to include Hamiltonian terms (\ref{Ht}), (\ref{Hg-t2}) and (\ref{Hf-t}),\begin{equation}
H=H_{g}+H_{f}+H_{V}+H_{U}+H_{t}+H_{g-t}+H_{f-t}\,.\label{eq:Hfull}\end{equation}
 In the following, we will assume perturbation theory in the hybridization
of the tip with the rest of the system, namely $t_{f}$ and $t_{x}$
($x=a,\, b$) are small compared to the hybridization $V$ of the
adatom with the host (graphene). This is not nevertheless a strict
requirement at the mean field level, and the 'exact' expressions of
the Green's functions are shown in the Appendix. In addition, we will
also assume that the system remains in equilibrium in the presence
of the STM tip. A detailed calculation of the equilibrium and also
out of equilibrium Green's functions in the STM problem for metallic
surfaces can be found in Ref. \onlinecite{plihal}.

To further simplify matters, we assume here that $\kappa_{\mathbf{p}}\sim\kappa$
is computed at the Fermi energy and hence is momentum independent,
in which case \begin{equation}
t_{x,\mathbf{k}\mathbf{p}}(\mathbf{r})\to t_{x}(z)\,\mbox{e}^{-i\mathbf{k}\cdot\mathbf{R}}\equiv t_{x,\mathbf{k}}(\mathbf{r})\,,\label{eq:t2}\end{equation}
 with $x=a,\, b$. Since the STM tip is typically large compared to
the lattice spacing in graphene, one can further simplify things by
assuming $t_{a,\mathbf{p}}=t_{b,\mathbf{p}}$. We will keep the $a,\, b$
labels below for completeness. A similar assumption will be made for
the tunneling matrix element between the tip and the adatom, $t_{f,\mathbf{p}}(r)\to t_{f}(r)$.

The matrix elements of the renormalized Green's function is the $a,\, b$
sublattice basis can be calculated straightforwardly,\begin{widetext}

\begin{eqnarray}
G_{xy,\sigma}(\mathbf{p},\mathbf{p}^{\prime},i\omega) & = & \delta_{\mathbf{p},\mathbf{p}^{\prime}}G_{xy}^{0}(p)+\Gamma_{x}(p,\mathbf{r})\Lambda_{x}(p)G_{ff,\sigma}(i\omega)\bar{\Lambda}_{y}(p^{\prime})\bar{\Gamma}_{y}(p^{\prime},\mathbf{r})+T_{x}(p,\mathbf{r})\bar{T}_{y}(p^{\prime},\mathbf{r})\sum_{\mathbf{k}}G_{cc}^{0}(k)\,,\label{eq:Gpp}\end{eqnarray}
 \end{widetext}The quantities $\Lambda_{x}(p)$ and $\bar{\Lambda}_{x}(p)$
were defined in Eq. (\ref{eq:Lambda2}) and (\ref{Lambdabar}), whereas
\begin{eqnarray}
T_{x}(p,\mathbf{r}) & \equiv & \sum_{y=a,b}G_{xy}^{0}(p)t_{y,\mathbf{p}}(\mathbf{r})\label{eq:thetaHa}\\
\bar{T}_{x}(p,\mathbf{r}) & \equiv & \sum_{y=a,b}t_{y,\mathbf{p}}^{*}(\mathbf{r})G_{yx}^{0}(p)\label{ThetaH}\end{eqnarray}
 contain the tunneling amplitudes and phases for the electrons as
they hop between the tip and the $A,\, B$ sublattices. $G_{cc}^{0}(\mathbf{k},\tau)=-\langle T[c_{\mathbf{k}}(\tau)c_{\mathbf{k}}^{\dagger}(0)]\rangle$
is the bare Green's function of the electrons of the tip, \begin{equation}
G_{cc}^{0}(p)=\frac{1}{i\omega-\epsilon_{\mathbf{p}}}\,,\label{eq:Gp}\end{equation}
 while $\Gamma$ and its conjugate form $\bar{\Gamma}$ in Eq. (\ref{eq:Gpp})
define the vertex renormalization due to the presence of the tip,
\begin{eqnarray}
\Gamma_{x}(p,\mathbf{r}) & = & 1+\frac{T_{x}(p,\mathbf{r})}{\Lambda_{x}(p)}\bar{t}_{f}(\mathbf{r},\omega)\sum_{\mathbf{k}}G_{cc}^{0}(k)\label{eq:Gamma}\\
\bar{\Gamma}_{x}(p,\mathbf{r}) & = & 1+\frac{\bar{T}_{x}(p,\mathbf{r})}{\bar{\Lambda}_{x}(p)}t_{f}(\mathbf{r},\omega)\sum_{\mathbf{k}}G_{cc}^{0}(k)\,,\label{eq:Gammabar}\end{eqnarray}
 where the quantities \begin{eqnarray}
t_{f}(\mathbf{r},i\omega) & = & t_{f}(r)+\sum_{y=a,b}\bar{\Lambda}_{y}(-\mathbf{R},i\omega)t_{y}(z)\label{eq:V-1}\\
\bar{t}_{f}(\mathbf{r},i\omega) & = & t_{f}(r)+\sum_{y=a,b}t_{y}^{*}(z)\Lambda_{y}(\mathbf{R},i\omega),\label{eq:Vbar}\end{eqnarray}
 give the renormalized tunneling functions between the tip and the
adatom, whose bare form, $t_{f}(r)$, is defined in Eq. (\ref{tf})
. We also defined \begin{eqnarray*}
\Lambda_{x}(\mathbf{R}) & = & \sum_{\mathbf{k}}\Lambda(k)\,\mbox{e}^{i\mathbf{k}\cdot\mathbf{R}}\\
\bar{\Lambda}_{x}(\mathbf{R}) & = & \sum_{\mathbf{k}}\bar{\Lambda}(k)\,\mbox{e}^{i\mathbf{k}\cdot\mathbf{R}}\end{eqnarray*}
 as the Fourier transforms of $\Lambda(k)$ and $\bar{\Lambda}(k)$
{[}see Eq. (\ref{eq:Lambda2}) and (\ref{Lambdabar}){]}.

The self-energy correction to the localized electrons, $\Sigma_{ff}(\omega)$,
as given in Eq. (\ref{SEH0}), is also dressed by the proximity of
the STM tip and assumes the form \begin{equation}
\Sigma_{ff}(\mathbf{r},i\omega)=\Sigma_{ff}(i\omega)+\Sigma_{ff}^{(1)}(\mathbf{r},i\omega)\label{eq:Sigmadressed}\end{equation}
 where\begin{equation}
\Sigma_{ff}^{(1)}(\mathbf{r},i\omega)=t_{f}(\mathbf{r})\bar{t}_{f}(\mathbf{r})\sum_{\mathbf{k}}G_{cc}(k)\,.\label{eq:SigRen}\end{equation}
 gives the contribution from the tip to leading order in $t_{f}$
and $t_{x}$. In the presence of the STM tip, the Green's function
of the localized electrons depends explicitly on the distance between
the tip to the adatom, \begin{equation}
G_{ff,\sigma}^{R}(i\omega)=\left[i\omega-\epsilon_{\sigma}-\Sigma_{ff}(\mathbf{r},i\omega)\right]^{-1}\,,\label{eq:GggTip}\end{equation}
which reflects the influence of the tip into the wavefunction of the
localized states.

The imaginary part of the self energy, $\mbox{Im}\Sigma_{ff}^{(1)}(\mathbf{r},i\omega)$,
renormalizes the level broadening $\Delta(\omega)$, defined in Eq.
(\ref{eq:Delta}), due to the hybridization of the localized electrons
with the electrons in the tip. In contrast with metallic hosts, which
have a large DOS, in graphene the metallic tip can locally overwhelm
the hybridization of the adatom with the nearby carbon atoms. In the
situation where the level broadening becomes large enough as to overcome
local correlation effects in the localized state, the tip might eventually
lead to suppression of the local magnetism. This effect will be discussed
in more detail in sec. IV.B.

Finally, other useful quantities are two off diagonal Green's functions
$G_{cx,\sigma}(\mathbf{p},\tau)=-\langle T[c(\tau)x^{\dagger}(0)]\rangle$,
with $x=a,b$, which are given by \begin{eqnarray}
G_{cx,\sigma}(\mathbf{p},\mathbf{p}^{\prime},i\omega) & = & G_{cc}^{0}(p)\left[T_{y}^{*}(p^{\prime},\mathbf{r})+\bar{t}_{f}(\mathbf{r},i\omega)\bar{\Gamma}_{x}(p^{\prime},\mathbf{r})\right.\nonumber \\
 &  & \qquad\qquad\left.\times G_{ff,\sigma}(i\omega)\bar{\Lambda}_{x}(p^{\prime})\right]\!,\label{eq:Gcx}\end{eqnarray}
and also\begin{equation}
G_{cf,\sigma}(p)=G_{cc}^{0}(p)\bar{t}_{f}(\mathbf{r},i\omega)G_{ff,\sigma}(i\omega)\,,\label{Gcf2prime2}\end{equation}
 which are required for computing the differential conductance (see
sec. IV. B.1).

\subsubsection{Local DOS}

Besides the localized state, the local DOS around the adatom is also
affected by the presence of the STM tip. The local DOS nearby the
impurity is also indirectly affected by the hybridization of the orbitals
of the tip with the adatom localized orbital. For instance, for a
magnetic adatom, the DOS is expected to be spin polarized on a given
sublattice $x$, \begin{equation}
\rho_{x,\sigma}(\mathbf{r},\omega)=-\frac{1}{\pi}\mbox{Im}\sum_{\mathbf{p},\mathbf{p}^{\prime}}G_{xx,\sigma}^{R}(\mathbf{p},\mathbf{p}^{\prime},\omega)\mbox{e}^{i(\mathbf{p}-\mathbf{p}^{\prime})\cdot\mathbf{R}}\,,\label{eq:rho0}\end{equation}
 where the diagonal Green's function $G_{xx,\sigma}(\mathbf{p},\mathbf{p}^{\prime},i\omega)$
is explicitly shown in Eq. (\ref{eq:Gpp}). In a more explicit form,\begin{widetext}
\begin{eqnarray}
\rho_{x,\sigma}(\mathbf{r},\omega) & = & -\frac{1}{\pi}\mbox{Im}\left[\sum_{\mathbf{p}}G_{xx}^{0}(p)+\left[\Gamma_{x}*\Lambda_{x}\right]\!(\mathbf{r})\, G_{ff,\sigma}(\omega)\,[\bar{\Gamma}_{x}*\bar{\Lambda}_{x}](\mathbf{r})+T_{x}(z)\bar{T}_{x}(z)\sum_{\mathbf{k}}G_{cc}^{0}(k)\right]\label{eq:rho}\end{eqnarray}
 \end{widetext}gives the local DOS per spin, where \begin{eqnarray}
\left[\Gamma_{x}*\Lambda_{x}\right]\!(\mathbf{r}) & \equiv & \sum_{\mathbf{p}}\mbox{e}^{i\mathbf{p}\cdot\mathbf{R}}\,\Gamma_{x}(p,\mathbf{r})\Lambda_{x}(p)\label{eq:conv}\\
{}[\bar{\Gamma}_{x}*\bar{\Lambda}_{x}](\mathbf{r}) & \equiv & \sum_{\mathbf{p}}\mbox{e}^{-i\mathbf{p}\cdot\mathbf{R}}\,\bar{\Gamma}_{x}(p,\mathbf{r})\bar{\Lambda}_{x}(p)\label{eq:conv_bar}\end{eqnarray}
 is the Fourier transform convoluted over the the product of the $\Gamma(p,\mathbf{r})$
and $\Lambda_{x}(p)$ functions (and their respective conjugate forms),
as defined in Eq. (\ref{eq:Lambda2}), (\ref{Lambdabar}) and (\ref{eq:Gamma}),
(\ref{eq:Gammabar}), while $T_{x}(z,\omega)$ is by definition $\left.T(\mathbf{r},\omega)\right|_{\mathbf{R}=0}$
{[}see Eq. (\ref{eq:thetaHa}){]}, and hence independent of the horizontal
distance between the tip and the adatom.

\begin{figure}[b]
\begin{centering}
\includegraphics[scale=0.33]{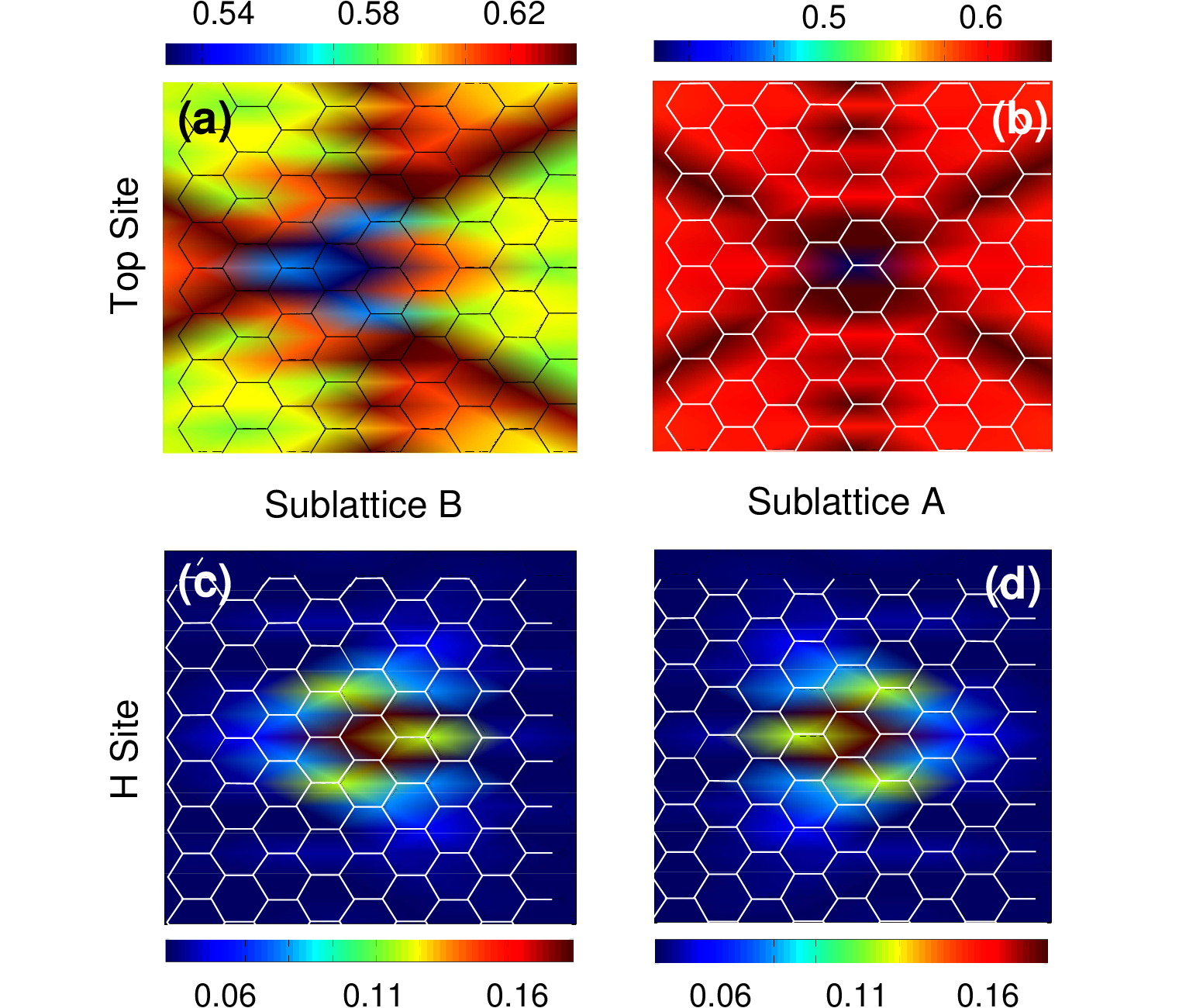}
\par\end{centering}

\caption{{\small Energy integrated LDOS around the adatom (center) in the presence
of an STM tip. a) Scans for the same sublattice of the impurity and
b) for the opposite sublattice (top carbon site case). Scans for sublattice
c) $A$ and d) $B$, nearby an adatom sitting at the center of a honeycomb
hexagon for some magnetic $m=0$ angular momentum state (center).
\label{fig:Integrated-LDOS-Tip}}}
\end{figure}

In Fig. \ref{fig:Integrated-LDOS-Tip} we show the topography maps
for LDOS in the presence of the STM tip for both sublattices in the
case of an adatom sitting on top of a carbon atom (top panels) and
also for an adatom on an $H$ site (lower panels). In those plots,
we use the same set of parameters as before, $V=1$eV, $U=1$eV, $\mu=0.1$eV,
$\epsilon_{0}=-0.5$eV, and additionally the parameters $\alpha_{D}=4$
eV for the band width of the tip and $\epsilon_{D}=2$eV for the Fermi
energy of the tip. The tunneling parameters between the tip and the
system where chosen to be $t_{f}=0.02$ eV and $t_{a}=t_{b}=0.2$eV.
When the STM tip is weakly coupled to the impurity, the plots show
basically the same qualitative features as the ones shown in Fig.
\ref{fig:Integrated-LDOS-around} for the actual DOS on graphene in
the absence of the STM tip. Due to the small DOS in the bath, additional
features reflecting the suppression of the local moment are possible
when the tip gets close to the adatom \cite{Uchoa2}. 

\subsection{Tunneling current}

The tunneling current from the tip is defined by\cite{plihal}:\begin{equation}
I=-e\left\langle \frac{\mbox{d}\hat{N}_{c}(t)}{\mbox{d}t}\right\rangle ,\label{I}\end{equation}
 where $\hat{N}_{c}=\sum_{\mathbf{k}\sigma}c_{\mathbf{k}\sigma}^{\dagger}c_{\mathbf{k}\sigma}$
is the number operator for the $c$ electrons in the tip, and $e$
is the electron charge. The motion equation for this operator is\begin{eqnarray*}
\partial_{t}\hat{N}_{c} & = & i\left[H,\hat{N}_{c}\right],\end{eqnarray*}
 where $H$ is the full Hamiltonian defined in Eq. (\ref{eq:Hfull})
including hopping matrix elements between the tip and the system.
After a straightforward algebra, the total current follows from the
sum of three different contributions that arise from tip tunneling
processes to either sublattices $A$, and $B$ or else to the adatom
localized state, \begin{eqnarray}
I & = & -2e\,\mbox{Im}\int_{-\infty}^{\infty}\frac{\mbox{d}\omega}{2\pi}\left[t_{f}(\mathbf{r})\sum_{\mathbf{k},\sigma}iG_{cf,\sigma}^{<}(\mathbf{k},\omega)\right.\nonumber \\
 &  & \qquad\qquad\left.+\sum_{\mathbf{k}\mathbf{p},\sigma}\sum_{x=a,b}t_{x,\mathbf{p}}(z)iG_{cx,\sigma}^{<}(\mathbf{k},\mathbf{p},\omega)\right],\quad\end{eqnarray}
 where\begin{eqnarray}
G^{<}(t,t^{\prime}) & \equiv & i\left\langle \psi^{\dagger}(t^{\prime})\psi(t)\right\rangle ,\label{G>}\\
G^{>}(t,t^{\prime}) & \equiv & -i\left\langle \psi(t)\psi^{\dagger}(t^{\prime})\right\rangle ,\label{G<}\end{eqnarray}
 are real time {}``lesser'' and {}``greater'' Green's functions,
which should be distinguished from retarded ($G^{R}$) and advanced
($G^{A}$) ones, which are time ordered. If $A(\omega)=B(\omega)C(\omega)$,
one may show that\cite{Kadanoff}\begin{equation}
A^{<}(\omega)=B^{R}(\omega)C^{<}(\omega)+B^{<}(\omega)C^{A}(\omega)\,,\label{A<}\end{equation}
 in which case the total current can be written as\begin{eqnarray}
I & = & -2e\,\mbox{Im}\int_{-\infty}^{\infty}\frac{\mbox{d}\omega}{2\pi i}\times\nonumber \\
 &  & \sum_{\mathbf{k}}\left[G_{cc}^{0\, R}(k)\Sigma_{cc}^{<}(\mathbf{r},\omega)+G_{cc}^{0\,<}(k)\Sigma_{cc}^{A}(\mathbf{r},\omega)\right]\label{eq:I3}\end{eqnarray}
 where\begin{eqnarray}
\Sigma_{cc}(\mathbf{r},i\omega) & = & \sum_{\mathbf{p}}\sum_{x=a,b}t_{x}(z)\bar{T}_{x}(z,p)\nonumber \\
 &  & \qquad+t_{f}(\mathbf{r})\bar{t}_{f}(\mathbf{r})\sum_{\sigma}G_{ff,\sigma}(i\omega)\label{eq:Sigma_cc}\end{eqnarray}
 is the self energy correction to the Green's function of the $c$-electrons
in the tip, \begin{equation}
G_{cc}(\mathbf{p},\mathbf{p}^{\prime},i\omega)=\left[(i\omega-\epsilon_{\mathbf{p}})\delta_{\mathbf{p},\mathbf{p}^{\prime}}-\Sigma_{cc}(\mathbf{r},i\omega)\right]^{-1}.\label{eq:Gcc}\end{equation}
 The first term in Eq. (\ref{eq:Sigma_cc}) gives the self-energy
contribution due to the graphene electrons, while the second one is
the contribution from the adatom. Using the fluctuation dissipation
theorem\cite{Kadanoff} \begin{equation}
G^{<}(\omega)=if(\omega)A(\omega)\,,\label{dissp-Fluct}\end{equation}
 where $A(\omega)=-2\mbox{Im}G^{R}(\omega)$ is the spectral function,
the total current is given by \begin{equation}
I(\mathbf{r},\omega^{\prime})=2\pi e\, t_{a}^{2}(z)\!\int_{-\infty}^{\infty}\mbox{d}\omega\,\rho_{t}(\mathbf{r},\omega)\rho_{c}(\omega^{\prime})\left[f(\omega)-f(\omega^{\prime})\right],\label{I5}\end{equation}
 where $f(\omega)=[\mbox{e}^{\omega/T}+1]^{-1}$ is the Fermi distribution,
$\rho_{c}$ is the DOS at the STM tip,\begin{equation}
\rho_{c}(\omega)=-\frac{1}{\pi}\sum_{\mathbf{k}}\mbox{Im}G_{cc}^{0}(k)\,,\label{rhoc}\end{equation}
 and $\rho_{t}$ is defined as \begin{equation}
\rho_{t}(\mathbf{r},\omega)=-\frac{1}{\pi t_{a}^{2}(z)}\mbox{Im}\Sigma_{cc}^{R}(\mathbf{r},\omega)\,.\label{rhot}\end{equation}
 This term has units of DOS, and accounts for the phases acquired
by the electrons in the tunneling process between the tip and the
localized state of the adatom. It can be conveniently rewritten in
the following form: \begin{eqnarray*}
\rho_{t}(\mathbf{r},\omega)\negthickspace & = & \negthickspace\rho_{0}(\omega)\left\{ \nu(\omega)+\frac{\pi}{2}\rho_{0}(\omega)V^{2}\right.\times\\
 &  & \negthickspace\left.\sum_{\sigma}\left[\left(\gamma\bar{\gamma}-q\bar{q}\right)\mbox{Im}G_{ff,\sigma}+(q\bar{\gamma}+\bar{q}\gamma)\mbox{Re}G_{ff,\sigma}\right]\right\} ,\end{eqnarray*}
 where \begin{equation}
\nu(\omega)=-\frac{1}{\pi V\rho_{0}(\omega)}\,\mbox{Im}\!\sum_{x=a,b}\bar{T}_{x}(\mathbf{R}=0,\omega)\,.\label{eq:nu}\end{equation}
 $\rho_{0}(\omega)$ is the bare local DOS of graphene in the absence
of the impurity and the tip, and $V\equiv\mbox{max}\left(V_{x,i}\right)$,
with $V_{x,i}$ the hybridization amplitudes of the adatom with the
nearest carbon atoms, as defined in Eq. (\ref{eq:Theta}). The other
parameter, $q$ and its conjugate form, $\bar{q}$ are the Fano factors,
\begin{eqnarray}
q(\mathbf{r},\omega) & = & \frac{1}{t_{a}(z)V}\,\frac{\mbox{Re}\, t_{f}(\mathbf{r},\omega)}{\pi\rho_{0}(\omega)}\label{qH}\\
\bar{q}(\mathbf{r},\omega) & = & \frac{1}{t_{a}(z)V}\,\frac{\mbox{Re}\,\bar{t}_{f}(\mathbf{r},\omega)}{\pi\rho_{0}(\omega)}\,,\end{eqnarray}
 while \begin{eqnarray}
\gamma(\mathbf{r},\omega) & = & -\frac{1}{\pi V\rho_{0}(\omega)}\sum_{x=a,b}\mbox{Im}\bar{\Lambda}_{x}(\mathbf{R},\omega)\label{eq:dump}\\
\bar{\gamma}(\mathbf{r},\omega) & = & -\frac{1}{\pi V\rho_{0}(\omega)}\sum_{x=a,b}\mbox{Im}\Lambda_{x}(\mathbf{R},\omega)\label{eq:dumpbar}\end{eqnarray}
 gives the corresponding damping factor. These factors characterize
the Fano resonances in the differential conductance curves in the
vicinity of a localized state.

\subsubsection{Differential conductance}

The Green's function of the localized electrons can be written in
a more compact form as:\begin{eqnarray}
G_{ff,\sigma}^{R}(\omega) & = & \frac{\xi_{\sigma}-i}{\xi_{\sigma}^{2}+1}\frac{1}{\mbox{Im}\Sigma_{ff}(\mathbf{r},\omega)}\,\label{eq:G3}\end{eqnarray}
 where $\Sigma_{ff}(\mathbf{r},\omega)$ is the dressed self-energy
of the localized electrons due to the proximity of the tip, as defined
in Eq. (\ref{eq:Sigmadressed}), and $\xi_{\sigma}(\omega)$ is defined
as \begin{equation}
\xi_{\sigma}(\mathbf{r},\omega)=\frac{\omega-\epsilon_{\sigma}-\mbox{Re}\Sigma_{ff}(\mathbf{r},\omega)}{\mbox{Im}\Sigma_{ff}(\mathbf{r},\omega)}\,.\label{xi4}\end{equation}
 The differential conductance follows by computing $I/V_{b}$ in the
limit of $V_{b}\equiv\omega^{\prime}-\omega\to0$. Since $\mbox{d}f(\omega^{\prime})/\mbox{d}\omega=-\delta(\omega^{\prime})$
at zero temperature, the differential conductance can be written in
the more standard form

\begin{eqnarray}
\mathcal{G}(\mathbf{R},\omega_{b})\! & =\! & 2\pi e\rho_{c}(0)t_{f}^{2}(z)\rho_{0}(\omega_{b})\times\label{dIdV3}\\
 &  & \!\sum_{\sigma}\left\{ \nu(\omega_{b})+\left[\frac{q\bar{q}-\gamma\bar{\gamma}+(q\bar{\gamma}+\bar{q}\gamma)\xi_{\sigma}}{\xi_{\sigma}^{2}+1}\right]\right\} ,\nonumber \end{eqnarray}
 where $\omega_{b}$ is the bias voltage. The first term in parenthesis
defines the DC due to the DOS in graphene. The second one is explicitly
defined in terms of the Fano parameters and gives the contribution
due to the presence of the magnetic adatom.

The experimental detection of a localized state with STM tips is based
on the principle of quantum interference between the two different
hybridization paths the electrons can take when they tunnel from the
impurity to the localized state. In one way, the electrons can tunnel
directly to the localized state. On the other, they can also tunnel
to the host material (graphene) and then hybridize with the localized
orbital. The signature of such interference appears in the differential
conductance curves in the form of a Fano resonance. In graphene, the
electrons have additional sublattice quantum numbers which may give
rise to additional interference effects, depending on the position
of the adatom relative to the two different sublattices. In the case
where the adatom sits in the center of the honeycomb hexagon, for
a given sublattice, there are three different paths the electrons
in graphene can take to hybridize with the adatom. Destructive interference
between the different paths in a given sublattice can suppress the
Fano character of the resonance and change the shape of the DC curves.

When the tip is above the adatom (\textbf{$R=0$}), the conjugate
forms $q=\bar{q}$ and $\gamma=\bar{\gamma}$ in Eq. (\ref{qH})-(\ref{eq:dumpbar})
are the same. In the simplest scenario, where an adatom sits on top
of a carbon atom, say on site $A$, the Fano factor is defined explicitly
in terms of the self-energy for orbitals of type I, $\mbox{Re}\Sigma_{ff}^{I}(\omega)=\omega[Z_{I}^{-1}(\omega)-1]$,
as given in Eq. (\ref{eq:Z1}), namely \begin{equation}
q^{A}(0,\omega)=\frac{V_{c}+(t_{a}(z)/V)\mbox{Re}\Sigma_{ff}^{I}(\omega)}{\pi t_{a}(z)V\rho_{0}(\omega)}\,.\label{eq:q}\end{equation}
 The damping in this case is $\gamma^{A}=1$, by noticing that integrals
with off diagonal matrix elements of the Green's function, such as
$\sum_{\mathbf{k}}G_{ab}=0$. In a different scenario, for adatoms
of type II, which sit either in $S$ or $H$ sites and possess orbitals
with $C_{3v}$ point group symmetry, as discussed in Sec. III, the
hybridization matrix elements have the form $V_{b,\mathbf{p}}=\pm V\phi_{\mathbf{p}}$
and $V_{a,\mathbf{p}}=V\phi_{\mathbf{p}}^{*}$, in which case one
can easily check that $\Lambda(\mathbf{R}=0,\omega)=\sum_{x,y=a,b}\sum_{\mathbf{k}}G_{xy}V_{y,\mathbf{p}}=0$.
In that case, \begin{equation}
q^{II}(0,\omega)=V_{c}/\left[\pi t_{a}(z)V\rho_{0}(\omega)\right],\label{eq:qII}\end{equation}
 and $\gamma^{II}=0$. In the more generic case, for type I orbitals
(the ones which are not $C_{3v}$ invariant) that sit on $H$ or $S$
sites, the damping factor $\gamma$ interpolates between 0 and $N_{s}=1,2$,
the number of sublattices the adatom effectively hybridizes. 

\begin{figure}
\begin{centering}
\includegraphics[scale=0.4]{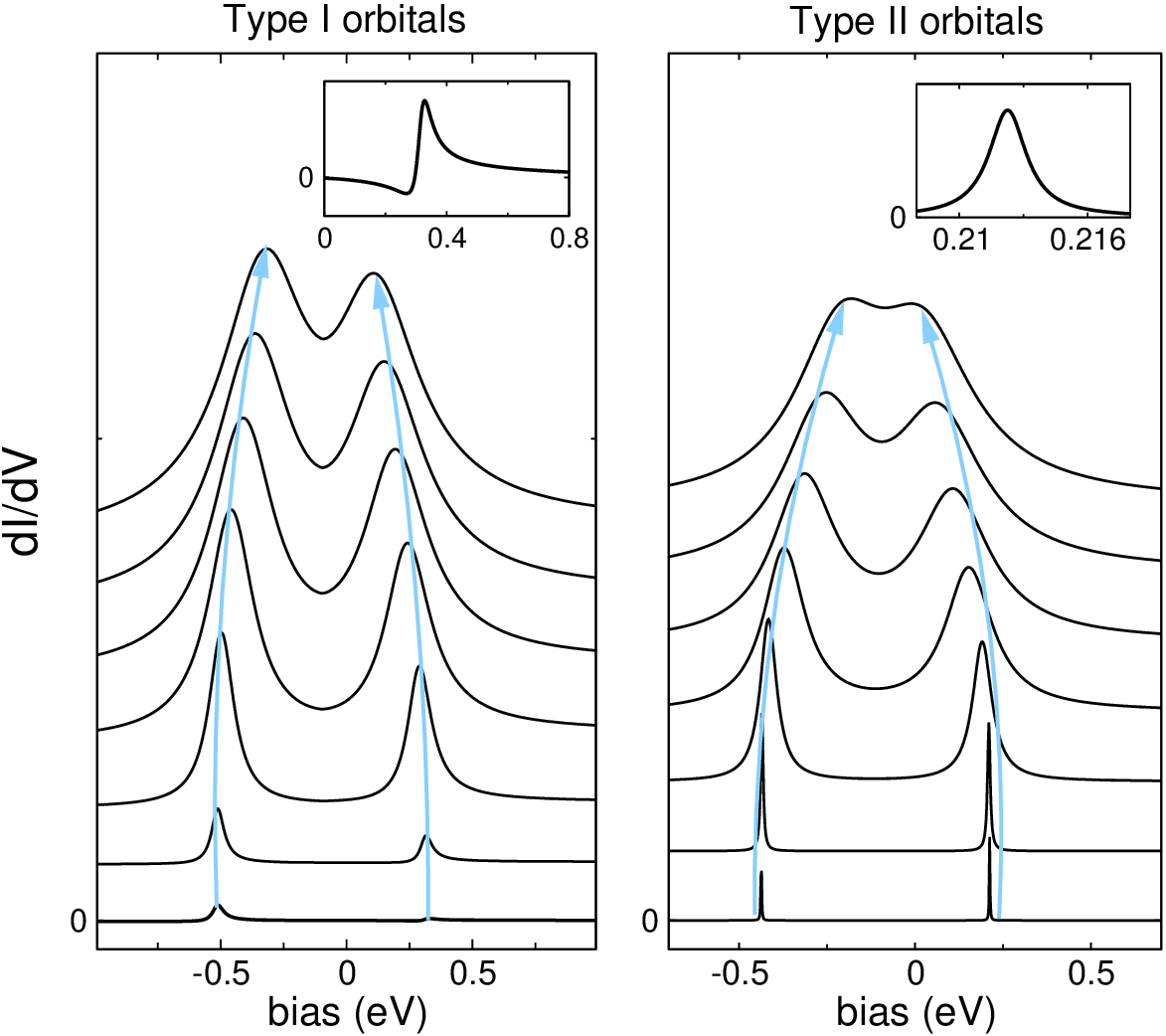} 
\par\end{centering}

\caption{{\small Differential conductance induced by the adatom versus bias
when the tip is right above the adatom ($\mathbf{R}=0)$ for $m=0$
states. The different curves are shifted vertically. Left: type one
orbitals, for adatoms that sit on top of carbon site; Right: type
II orbitals at $S$ or $H$ sites ($m=0)$. See details in the text.
a), b) $t_{a}=t_{b}=0.15$eV and $t_{f}/t_{a}=1.6$, $1.5$, $1.35$,
$1.1$, $0.7$, $0.25$ and $0.1$ (inset), from top to bottom.\label{fig:Adatom-induced-DC,}}}
\end{figure}

As the usual theory of Fano resonances\cite{fano}, when $q/\gamma\gg1$,
the DC curve has the form of a peak, whereas in the opposite limit,
for $q/\gamma\ll1$ it is shows a dip. In Fig. \ref{fig:Adatom-induced-DC,}
we show the DC induced by the presence of the adatom for $m=0$ orbitals
in two cases: when it sits on top of a carbon site (left panels) and
also when it sits at an $H$ site (right panels). In panels a) and
b) we assume a fixed set of parameters and change the ratios $t_{f}/t_{a}$
for a given fixed value of $t_{a}=0.15$eV. For $t_{f}/t_{a}\lesssim0.1$,
the DC curve of type I orbitals has a Fano shape, whereas for all
type II orbitals the Fano resonance is suppressed (see insets of Fig.
\ref{fig:Adatom-induced-DC,}).

The evolution of the separation of the peaks can be rigorously analyzed
within the single orbital model for the case of $m=0$ orbitals. The
increase of $t_{f}$ leads to a gradual suppression of the local magnetic
moment, and as a consequence to a decrease in the separation of the
two peaks. In the $m\neq0$ case, a significant suppression of the
local moment by the STM tip is also accompanied by a redistribution
of the charge between different orbitals contained in given irreducible
representation.

\section{Conclusion}

Unlike the case of metallic hosts, in graphene the symmetry of the
localized orbital has clear fingerprints in the LDOS nearby the adatom
whenever the adatom, hybridizes with two or more carbon atoms. We
showed that the real and momentum space STM scanning maps can reveal
not only the position of the adatom with respect to the sublattices
but can also indicate the orbital symmetry of the localized state
and possibly its magnetic state.

We have described in detail how sublattice quantum numbers in combination
with orbital symmetry effects influence the Fano resonances in the
differential conductance nearby the adatom. To illustrate the effect,
we analyzed the problem in the single orbital picture, which is valid
for orbitals in one dimensional irreducible representations (such
as $s$-wave, $d_{zz}$wave, etc), as well as in the more interesting
case of 2 dimensional irreducible representations, such as in the
doublets ($d_{x^{2}-y^{2}},d_{xy})$, $(f_{x(x^{2}-3y^{2})},f_{y(3y^{2}-x^{2})}),$
etc, when the energy separation of the orbitals in the doublet state,
driven for instance by charge and spin polarization effects, is large
compared to the level broadening.\textcolor{blue}{{} }\textcolor{black}{In
the presence of small Jahn -Teller distortion effects, which freeze
charge fluctuations between the polarized orbitals and hence break
the point group symmetry of the underlying crystal, those different
orbitals may leave explicit fingerprints in the LDOS. Those distortions
can occur either spontaneously, when adatoms are randomly adsorbed
on graphene, or else through the application of strain. }

We acknowledge P. M. Goldbart, E. Andrei, H. Manoharan, A. Balatsky,
K. Sengupta and M. Vojta for discussions. BU acknowledges partial
support from DOI grant DE-FG02-07ER46453 at the University of Illinois.
LY and SWT acknowledge financial support from NSF under grant DMR0847801
and from the UC-Lab FRP under award number 09-LR-05-118602. AHCN acknowledges
DOE grant DE-FG02-08ER46512 and NRF-CRP award \textquotedbl{}Novel
2D materials with tailored properties\textquotedbl{} beyond graphene\textquotedbl{}
(R-144-000-295-281).

\appendix

\section{Exact Green's functions}

At the mean field level, the Green's functions described in sec. IV
can be written in an exact form by solving the equations of motion
for the fermionic operators $a_{\mathbf{k},\sigma}$, $b_{\mathbf{k},\sigma}$,
$c_{\mathbf{p},\sigma}$ and $f_{\sigma}$. After a cumbersome but
straightforward algebra, the final expressions are,\begin{widetext}

\begin{equation}
G_{xy,\sigma}(\mathbf{p},\mathbf{p}^{\prime},i\omega)=\delta_{\mathbf{p},\mathbf{p}^{\prime}}G_{xy}^{0}(p)+\Gamma_{x}(p,\mathbf{r})\Lambda_{x}(p)G_{ff,\sigma}(i\omega)\bar{\Lambda}_{y}(p^{\prime})\bar{\Gamma}_{y}(p^{\prime},\mathbf{r})+\sum_{\mathbf{k}\mathbf{k}^{\prime}}T_{x,\mathbf{k}^{\prime}}(\mathbf{r},p)G_{cc}(\mathbf{k},\mathbf{k}^{\prime},i\omega)\bar{T}_{y,\mathbf{k}^{\prime}}(\mathbf{r},p^{\prime})\,,\label{eq:GxyT}\end{equation}
 \begin{equation}
G_{cx,\sigma}(\mathbf{p},\mathbf{p}^{\prime},i\omega)=\sum_{\mathbf{k}}G_{cc}(\mathbf{p},\mathbf{k}^{\prime},i\omega)\left[\bar{T}_{x,\mathbf{k}}(\mathbf{r},p^{\prime})+\bar{t}_{f,\mathbf{k}}\bar{\Gamma}_{x,\mathbf{p}^{\prime}}G_{ff,\sigma}(i\omega)\bar{\Lambda}_{x}(p^{\prime})\right],\label{eq:GcxT}\end{equation}
 \end{widetext}and\begin{equation}
G_{cf,\sigma}(\mathbf{p}^{\prime},i\omega)=\sum_{\mathbf{p}}G_{cc}(\mathbf{p},\mathbf{p}^{\prime},i\omega)\bar{t}_{f,\mathbf{p}}(\mathbf{r})G_{ff,\sigma}(i\omega)\,,\label{Gcf2prime2}\end{equation}
 where $p\equiv(\mathbf{p},i\omega)$. The quantities $\Lambda_{x}(p)$
and $\bar{\Lambda}_{x}(p)$ have the same definitions as before {[}see
Eq. (\ref{eq:Lambda2}) and (\ref{Lambdabar}{]}, whereas\begin{eqnarray}
T_{\mathbf{p}^{\prime},x}(\mathbf{r},p) & \equiv & \sum_{y=a,b}G_{xy}^{0}(p)t_{y,\mathbf{p},\mathbf{p}^{\prime}}(\mathbf{r})\label{eq:thetaHaA}\\
\bar{T}_{\mathbf{p}^{\prime},x}(\mathbf{r},p) & \equiv & \sum_{y=a,b}t_{y,\mathbf{p},\mathbf{p}^{\prime}}^{*}(\mathbf{r})G_{yx}^{0}(p)\,.\label{ThetaHA}\end{eqnarray}
 The Green's function of the $c$-electrons is defined as \begin{equation}
G_{cc}(\mathbf{p},\mathbf{p}^{\prime},i\omega)=\left[(i\omega-\epsilon_{\mathbf{p}})\delta_{\mathbf{p}\mathbf{p}^{\prime}}-\Sigma_{cc,\mathbf{p}\mathbf{p^{\prime}}}(z,i\omega)\right]^{-1}\,,\label{eq:Gcc5}\end{equation}
 where $\Sigma_{cc}$ is the self-energy for the electrons in the
tip due to hybridization effects with the electrons in graphene only,\begin{equation}
\Sigma_{cc,\mathbf{p}\mathbf{p}^{\prime}}(z,i\omega)=\sum_{\mathbf{k}}\sum_{x=a,b}t_{x,\mathbf{p}}(z)\bar{T}_{\mathbf{p}^{\prime},x}(z,k)\,.\label{eq:Sigmacc4}\end{equation}
 The Green's function of the $f$-electrons is given by\begin{equation}
G_{ff,\sigma}(i\omega)=\left[i\omega-\epsilon_{\sigma}-\Sigma_{ff}(\mathbf{r},i\omega)\right]^{-1}\,,\label{eq:Gff4}\end{equation}
 where $\Sigma_{ff}(\mathbf{r},i\omega)=\Sigma_{ff}(i\omega)+\Sigma_{ff}^{(t)}(\mathbf{r},i\omega)$
is the corresponding self-energy, with\begin{equation}
\Sigma_{ff}^{(t)}(\mathbf{r},i\omega)=\sum_{\mathbf{k}\mathbf{k}^{\prime}}t_{f,\mathbf{k}}(\mathbf{r})G_{cc}(\mathbf{k},\mathbf{k}^{\prime},i\omega)\bar{t}_{f\mathbf{k}^{\prime}}(\mathbf{r})\label{eq:SigRen}\end{equation}
 as the contribution of the tip. The other quantities include the
renormalized hybridization of the tip with the adatom,\begin{eqnarray}
t_{f,\mathbf{p}}(\mathbf{r},i\omega) & = & t_{f,\mathbf{p}}(r)+\sum_{\mathbf{k}}\sum_{y=a,b}\bar{\Lambda}_{y}(k)t_{y,\mathbf{k}\mathbf{p}}\label{eq:tf4}\\
\bar{t}_{f,\mathbf{p}}(\mathbf{r},i\omega) & = & t_{f,\mathbf{p}}(r)+\sum_{\mathbf{k}}\sum_{y=a,b}t_{y,\mathbf{k}\mathbf{p}}^{*}\Lambda_{y}(k)\,,\qquad\label{eq:tf4bar}\end{eqnarray}
 and\begin{eqnarray*}
\Gamma_{x}(p) & \negthickspace\negthickspace=\negthickspace & 1+\sum_{\mathbf{k}\mathbf{k}^{\prime}}\bar{t}_{f,\mathbf{k}}(\mathbf{r},i\omega)G_{cc}(\mathbf{k},\mathbf{k}^{\prime},i\omega)\frac{T_{\mathbf{k}^{\prime},x}(\mathbf{r},p)}{\Lambda_{x}(p)}\\
\bar{\Gamma}_{x}(p) & = & 1+\sum_{\mathbf{k}\mathbf{k}^{\prime}}\frac{\bar{T}_{\mathbf{k}^{\prime},x}(\mathbf{r},p)}{\bar{\Lambda}_{x}(p)}G_{cc}(\mathbf{k}.\mathbf{k}^{\prime},i\omega)t_{f,\mathbf{k}}(\mathbf{r},i\omega),\end{eqnarray*}
 which are vertex corrections that appear in Eq. (\ref{eq:GxyT})
and (\ref{eq:GcxT}).

\end{document}